\documentclass[pre,amsmath,amssymb,10pt,nofootinbib,showpacs]{revtex4}

\usepackage{graphicx}
\usepackage{dcolumn}
\usepackage{bm}

\def\eref#1{Eq.~\ref{#1}}
\def\erefs#1{Eqs.~\ref{#1}} 
\def\sref#1{Sect.~\ref{#1}}
\def\fref#1{Fig.~\ref{#1}}
\def\rref#1{Ref.~\cite{#1}}
\def\rrefs#1{Refs.~\cite{#1}}
\def\aref#1{Appendix Sect..~\ref{#1}}
\def\SEC{SEC}

\bibstyle{plain}

\newcommand{\pnalt}[2]{{{\footnotesize{\it{#1}}}\bf{#2}}}

\newcommand{\pfign}[4]{
\begin{figure}
\begin{center}
\includegraphics{#1}
\caption{\label{#2} #4}
\end{center}
\end{figure}
}

\def\aref#1{Appendix~\ref{#1}}
\def\rrefs#1{Refs.~\cite{#1}}

\def\pnlabel#1{\label{#1}}

\normalsize

\long\def\pnalt#1#2{{#1}}{}

\def\srefs#1{Sects.~\ref{#1}}
\def\tot{_{\rm tot}}
\def\that{\vec t}
\def\kbt{k_{\rm B}T}

\begin{document}


\title{A generalized theory of semiflexible polymers}

\author{Paul A. Wiggins}
\email{pwiggins@caltech.edu} \affiliation{Division of Physics,
Mathematics, \& Astronomy, California Institute of Technology,
Pasadena CA 91125, USA}
 \homepage{http://www.rpgroup.caltech.edu/~wiggins/}

%

\author{Philip C. Nelson}
\email{nelson@physics.upenn.edu} \affiliation{Department of
Physics and Astronomy, University of Pennsylvania, Philadelphia PA
19104, USA}

\date{\today}


\begin{abstract}
%
%
%
%
%
%
%
DNA bending on length scales shorter than a persistence length 
plays an integral role in the translation of genetic
information from DNA to cellular function. Quantitative experimental 
studies of these biological systems have led to a renewed interest in the polymer 
mechanics relevant for describing the conformational free energy of DNA bending 
induced by protein-DNA complexes. 
Recent experimental results from DNA cyclization studies have cast doubt on the 
applicability of the  canonical semiflexible polymer theory, the wormlike chain (WLC) model, to DNA bending on biological length scales.

This paper develops a theory of the chain statistics of a class
of generalized semiflexible polymer models. Our focus is on the theoretical
development of these models and the calculation of experimental
observables. 
To illustrate our methods, we focus on a specific toy model
of DNA bending.
We show that the WLC model generically describes the
long-length-scale chain statistics of semiflexible polymers, as predicted by
the Renormalization Group. In particular, we show that either the WLC 
or our new model adequate describes force-extension, solution scattering, and
long-contour-length cyclization experiments, regardless of the details of DNA bend 
elasticity.
In contrast, experiments sensitive to short-length-scale chain behavior 
can in principle reveal dramatic departures from the linear elastic behavior assumed in the WLC
model. We demonstrate this explicitly by showing
that our toy model can reproduce the anomalously large
short-contour-length cyclization $J$ factors observed by Cloutier and
Widom.  Finally, we discuss the applicability of these models to DNA
chain statistics in the context of future experiments.

\end{abstract}


\maketitle

\section{Introduction}

The statistical mechanics of linear polymers has long attracted the attention
of physicists and chemists alike. The mechanics of DNA is of considerable biological relevance to describing 
the free energy landscape controlling protein-induced DNA bending. These protein-DNA interactions are
of central importance to cellular function on a microscopic scale, from chromosomal DNA
packaging, to transcription, and gene regulation, to viral
packaging \cite{Alberts}. Protein-DNA interactions typically induce
short-length-scale DNA bending which couples the chemical and
physical properties of DNA \cite{widomnuclrev,Rippe1995,Shore81}.

A particularly important and successful application of polymer statistics has been 
in the description of double stranded DNA (dsDNA) by the wormlike chain model (WLC).
In the WLC model, DNA is modeled as a fluctuating, linearly-elastic rod.
This simple model has been remarkably successful in describing 
many aspects of DNA mechanics and the statistics of semiflexible polymers generally. 
In particular, WLC describes 
the extension of a single dsDNA molecule under an external force with impressive precision \cite{Bustamante2000}. 


Despite the notable theoretical and experimental success of the wormlike chain model, 
recent DNA cyclization  studies by Cloutier and Widom \cite{CW}
have cast doubt on the 
validity of the WLC model for describing the cyclization of short-contour-length sequences of DNA. 
In still more recent cyclization studies, Vologodskii and coworkers claim 
that the WLC model does accurately describe the cyclization of short
DNA sequences \cite{Du:2005ir}. Nevertheless, as we will explain
later a
number of experiments do seem to point to a role for elastic breakdown
in DNA mechanics.

With the current experimental
situation still in flux, it seems imperative
to reevaluate the WLC model {\it theoretically.} We wish
to answer the questions: {\it(i)} How could such a simple theory
hope to describe a complex molecule like DNA? {\it(ii)} More precisely, which
classes of experiments would we expect to be successfully described by
WLC model, and which might require a different
theory? Do these experiments correspond to the known successes or
the recently reported failures of the theory? 
In other words, we are asking how much {\it room} do the classic
tests of WLC model leave for generalization of this model, and how completely do
these experiments test the WLC model? Finally, we must ask
{\it(iii)} Would a breakdown of the WLC model have any biological significance? 

The focus of this paper will be the theoretical analysis of these questions and the development and discussion of more general semiflexible polymer models. Although these ideas are widely applicable to polymers statistics in general, the focus of this paper will be exclusively the mechanics of DNA. We shall attempt a synthesis of the existing experimental knowledge to determine which aspects of DNA bending are probed by existing experiments. In particular, we determine which experiments are most sensitive to the DNA mechanics relevant for understanding biological systems.   
In the remainder of this introduction, we shall quickly outline our answers to the questions posed above.

%
%

\subsection{Scale dependence in statistical physics}
First, to put the possible breakdown of the WLC model into perspective, it is helpful to consider the bending of macroscopic rods. To engineers in the mechanics community, whose work has been the study of
macroscopic bending, the failure of a linear elastic model at high curvature 
is more pedestrian than remarkable. The linear elastic theory is understood to 
apply only to the small deflection limit. What is perhaps more remarkable to some 
is that a linear-elastic model describes a macromolecular polymer at all, let alone 
to the accuracy illustrated by force-extension measurements!

To put the success of the WLC model into perspective, it is helpful to consider  
DNA mechanics from the viewpoint of the statistical mechanics of condensed matter systems. Many physical properties of complicated condensed matter systems have been described by a small set of theories described in terms of renormalizable operators \cite{Fisher1998}. 
Regardless of the complicated structure of the theory at short length scales, the 
Renormalization Group (RG) guarantees that the long-length-scale chain statistics will be described 
\pnalt{by an effective energy functional containing only a few terms. 
In fact, for semiflexible polymers, only one such ``renormalizable'' term
exists}{}{} with the right symmetries. As a consequence, all semiflexible polymers share generic long-length-scale behavior: that described by the WLC model.
Physically, this loss of information about the microscopic details of
molecular mechanics arises from the averaging effect of
thermal fluctuations. 

The RG world-view leads us to expect that experiments like measuring
the force-extension relation of long DNA would reveal only generic
behavior, insensitive to microscopic details of DNA elasticity.
But, on short enough length scales, the underlying structure of the theory becomes important. 
Violations of the linear elastic theory, analogous to those observed in macroscopic bending, are therefore expected in experiments that 
probe the short-length-scale bending of DNA. Indeed, early AFM imaging
experiments did see the onset of deviations from WLC expectations on
short scales \cite{Rivetti1996}.  Cyclization experiments, like the ones in 
\rrefs{CW,Du:2005ir}, hold the prospect of
greater sensitivity to the high-curvature regime.

\subsection{Summary of this paper}
This paper develops the qualitative framework outlined above by
introducing a generalization of the wormlike chain model.  This class
of models, introduced in \sref{gsdeftheor}, generalizes the WLC by 
describing a semiflexible polymer
by an arbitrary local bending energy function. 
\srefs{gSECM}--\ref{gFRET} introduce an explicit toy model of DNA
bending, the ``Sub-Elastic Chain'' model (SEC), motivated by imaging data on
DNA adsorbed
to mica \cite{VanNoort}, and by recent nanoscale force measurements \cite{Shroff2005}.  \sref{gprop} illustrates a computational
procedure for computing the tangent distribution function for
arbitrary contour length in generalized theories. \sref{gspersistence}
introduces the persistence length in generalized
theories and  shows that these theories
converge to the WLC model at long contour length.

The remainder of the paper focuses on the spatial distribution
of the polymer. The spatial distribution is of particular importance
for biological applications where the contribution of chain statistics
to biological function can often be formulated in terms of an
effective end-concentration, the Jacobson--Stockmayer factor ($J$
factor). Physically, this effective concentration is the
probability density of the polymer having the correct configuration
for binding to the binding site of a protein. \sref{gssdf} introduces
a method for computing the spatial and tangent-spatial
distributions of generalized semiflexible polymer models in terms of a
framework developed by Spakowitz and Wang \cite{Andy,Andy2,Andy3} and
others \cite{SS}. \sref{gtsdfs} explicitly computes the spatial
distributions for both the SEC and WLC models for various contour
lengths. We discuss the Renormalization Group applied to spatial
distributions and show the predicted convergence of the SEC and WLC
models at long contour length. \srefs{gfexts} and \ref{gssf} 
show that the force extension and the structure factor computed for
general theories are nearly identical to the WLC model results,
implying that these experiments do not probe the high-curvature chain
statistics important for many biological processes. \sref{gcycls}
computes the cyclization $J$ factor for generic theories. We show
that the SEC model gives rise to the enhanced cyclization efficiency
for short-contour-length sequences observed by Cloutier and Widom
\cite{CW} while leaving the long-contour-length $J$ factor identical
to that predicted by the WLC model.  Finally, we discuss the results
of this paper in the context of the recent cyclization measurements of
Vologodskii and coworkers \cite{Du:2005ir} and recent measurements of 
the deflection force for short sequences of DNA \cite{Shroff2005}.

\subsection{Relation to other recent work\label{s:rrw}}
Following Crick and Klug's initial suggestion that DNA might kink
\cite{Crick}, many classical works investigated the structural
implications of this conformational change at the single-basepair
level (reviewed in \cite{dick98a}). Indeed, many known DNA-protein complexes do display kinks in
the DNA backbone. In contrast, our focus here is on physical measurements of DNA
mechanics on a mesoscopic, several-basepair scale relevant for
biological processes like DNA looping in vivo.
As described in \sref{gSECM}, both Yan and Marko and we previously
formulated and solved ``kinking'' models, in which
DNA is assumed to undergo a sudden loss of bend stiffness beyond a
critical stress \cite{Yan2004,wiggins1,YanPRE}. Other related models were also formulated and solved
in \rrefs{popov04} and \cite{ranj05a}.
Sucato et al.\ have also performed Monte
Carlo simulations of kinkable chains, to obtain information about their structural and
thermodynamic properties \cite{suca04a}. Unlike these prior articles, the present work explores
the proposal that the breakdown of linear elasticity, when
coarsegrained to the mesoscopic scale, is effectively less abrupt than
in the kinking models. We suggest that such a model can reconcile the
growing evidence for elastic breakdown with the generic absence of
sharply kinked states when tightly-bent DNA is observed
microscopically.

\section{Defining discrete link theories\pnlabel{gsdeftheor}}
\subsection{Local energy functions\pnlabel{ss:lef}}
\pnalt{In this paper, we discuss a class of generalized elastic models for
the statistical mechanics of semiflexible, inextensible polymers.
The theories we discuss will be applicable to the description of
polymers on length scales longer than the scale of the molecular
structure. Accordingly, we idealize a semiflexible polymer as a chain (or
``rod'') consisting of $N$ discrete segments (``links''), each of
length $\ell$, joined by semiflexible hinges (``vertices;'' see
\fref{gdiscretechain}). The link length $\ell$ should be taken to be shorter than the scale of the experiment we wish to describe,
for example, shorter than the total length of the DNA in a cyclization
experiment.

We then introduce a coarse-grained free energy
cost for each chain configuration 
\begin{equation}\label{pnstar}
E=\sum_{j=1}^NE_j,
\end{equation}
where $E_j$ is the energy associated with vertex $j$.  To make a connection 
with the continuum mechanics picture, it is convenient to write this vertex bending energy as 
an energy density $\varepsilon$:
\begin{equation}
E_j=\ell\varepsilon(\ldots,\that_{j-1},\that_j,\ldots; j),
\end{equation}
that is a function of the $N$ tangent vectors $\{\that_1,\ldots,\that_N\}$ and the vertex number $j$.
The coarse-grained configurational free energy $E$ is a
combination of entropic and energetic parts, which depend on the
underlying molecular structure of the polymer. We will ignore the
effects of excluded volume, since we will be principally interested in
bending on length scales where  self-interaction
effects play a negligible role in describing the chain
conformation. We will also not allow for long-range interactions (that
is, longer than $\ell$); for instance, we assume that the solution
conditions fix an electrostatic screening length smaller than $\ell$.
\pfign{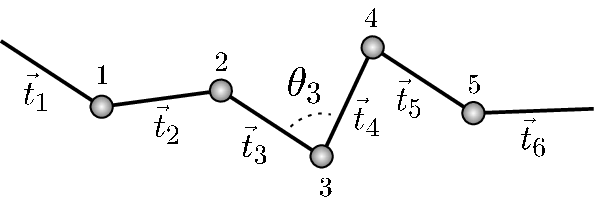}{gdiscretechain}{}{Link and vertex numbering. The energy is a function of the deflection angles. The deflection angle between links $i$ and $i+1$ is $\theta_i$. }

To focus our attention on the novel effects of the hypothesis of
elastic breakdown, we will restrict \eref{pnstar} to a subclass of
models by making some assumptions about the form of the free energy
density $\varepsilon$. Although this subclass is not a fully realistic
depiction of known properties of DNA, it does have the virtue of being
analytically tractable. Features of real DNA neglected in our models
can be introduced in more numerical approaches once the phenomena we
study are appreciated.
}{The bending energy density of a general model of the sort just
described can be written in terms of the arc length $s$, the
tangent vector $\vec t$ along the curve, and its derivatives: 
\begin{equation}
\varepsilon =
\varepsilon(\vec{t},d\vec{t}/ds,d^2\vec{t}/ds^2,...;s),
\end{equation}
where the energy density depends on arbitrarily high derivatives
of the tangent vector $\vec{t}$. We shall not study this completely
general framework, but a restricted subclass of models.}

First we shall assume that the free energy density does not explicitly
depend on the position $j$ (the chain is homogeneous). Strictly, this
is not the case for DNA since both the helical pitch and the sequence
dependence spoil homogeneity \cite{widomnuclrev,CW}.  We will
study the mechanics of DNA on length scales longer than the helical
repeat (3.6 nm), where helical effects approximately average
to zero. Sequence dependence is a more serious omission \cite{CW}, but
we make this approximation in order to get analytical formulas. Having
agreed to neglect helical effects, it is reasonable to add the assumption that the
theory is rotationally invariant (the bending stiffness is
isotropic).  Last, we assume that the energy density involves only the
first derivative of $\vec t$, the discretized version of curvature:
\begin{equation}
\vec{\kappa}(s) \equiv \Delta\vec{t}/\Delta s\quad\mbox{where\ }s_j=j\ell.
\end{equation}
By rotational invariance, the energy density is a function of the
magnitude of the curvature only: 
\begin{equation}\pnlabel{pne1}
\varepsilon =
\varepsilon(\kappa),\quad\mbox{where\ }\kappa_j=\theta_j/\ell \approx \|\vec{\kappa}\|.
\end{equation}

\pnalt{The most questionable of the assumptions above is the dropping
of higher derivatives, which we will call ``locality.'' For example, the
role of nonlinear elastic-energy terms such as $\kappa^4$ will be a
central concern of this paper, but this term has the same dimension
$(\mathrm{length})^{-4}$ as a term like $(\Delta\vec \kappa\,/\Delta s)^2$, which
we drop. Our justification is that higher-derivative terms correspond
physically to cooperative conformational changes along the polymer, and
although there are hints of long-range cooperativity in DNA
\cite{schur97a}, still the phenomena addressed in this paper do not seem
to require such behavior. Again, our restricted class of theories is an
analytically tractable starting point for the study of the effects of
nonlinear-elastic behavior in an entropy-dominated chain.}{
Of these assumptions, the most subtle is that associated with discarding the derivatives of the curvature. In the interest of brevity, we shall call this condition locality since it describes theories of semiflexible polymers with the fewest derivatives of the tangents.  It is not manifestly imperative that this restricted class of models describes DNA.  We have chosen this subset of general models motivated by Atomic Force Microscopy (AFM) experiments which we shall discuss in \sref{gSECM}.} We 
will return to this point again briefly in the Discussion (\sref{s:dis}). 

An important one-parameter family of polymer theories is described
by an energy density that is quadratic in the curvature:
\begin{equation}
\pnlabel{gWLCED}
 \varepsilon ={\textstyle \frac{1}{2}\xi} \kappa^2.
\end{equation}
\pnalt{The models described by \eref{gWLCED} have a
restoring torque $-d\varepsilon/d\theta$ that is linear in the link deflection $\theta$: they
are linear-elasticity theories.  Their continuum limits are called
wormlike chain models. }{} A WLC model is completely characterized by 
one number, the elastic bending modulus $\xi\kbt$. 
%


\subsection{Statistical mechanics\pnlabel{gintroSLC}}
We define the statistical mechanics theory associated to an
energy function $\varepsilon(s)$ in the canonical way. The probability
of a coarse-grained chain conformation is given by the Boltzmann law:
\begin{equation}
\pnlabel{gboltzmann}
{\cal P} = Z^{-1}\exp( -E),
\end{equation}
where $Z$ is the partition function, determined by normalization, and we
have defined the effective-coarse-grained free energy $E$ in units of $k_BT$.
\pnalt{}{ In fact, we shall simply define the energy by \eref{gboltzmann}. }

\pnalt{The link length $\ell$ plays two key roles in \eref{gboltzmann}.
First, it defines the coarse-grained configuration space: a chain of
physical length $L\tot$ consists of $N=L\tot/\ell$ links, each with
its orientation variable $\that_j$. Second,
$\ell$ enters \eref{pne1} explicitly. The physical meaning of $\ell$
is not obvious, however---it does not correspond to any
crystallographic length in the DNA structure, for example the basepair
rise of 0.34\thinspace{}nm. In fact, strictly speaking $\ell$ is not a
parameter of the theory at all, because two different values of $\ell$
can give rise to theories with identical predictions, if the two
theories' energy functions $\varepsilon$ are suitably adjusted.
Instead, $\ell$ is needed to give {\it meaning} to the energy function
$\varepsilon$. The
adjustment needed to maintain a fixed theory as $\ell$ is changed is
called the ``renormalization group flow'' of $\varepsilon$ \cite{Fisher1998}. 

It may seem tempting to eliminate $\ell$ from the theory by attempting
a continuum limit, $\ell\to0$, and indeed continuum mechanics does
just this. When fluctuations are important, however, the continuum
limit can discard some legitimate physics, and so must be treated with
caution. In fact, we will argue that in the polymer context, 
the continuum limit leads only to a subset of models corresponding to 
the WLC (\eref{gWLCED}), because
}{To define a statistical mechanics theory, the notion of an energy
density is not sufficient. We must also introduce a fundamental length
scale on which to define the theory. To determine the partition
function $Z$, we must have an explicit method for enumerating all
possible states.
To be explicit, we define the theory at scale $\ell$ by discretizing
the configuration into links length $\ell$, separated by torsional
springs whose energy is a function of the deflection angle $\theta_i$
between adjacent links $i$ and $i+1$. This discretized chain is
pictured schematically in \fref{gdiscretechain}.
We shall interpret (or define) the bending energy
density by taking it to be the bending energy for a uniformly bent
arc with deflection angle $\theta_i$ and length $\ell$:
\begin{equation}
E_i = \ell \varepsilon(\kappa_i),
\end{equation}
where $\kappa_i=\theta_i/\ell$.
The total bending energy $E$ for a configuration of links is the sum
of these vertex energies.

In mechanics, a continuum model would be defined by the limit as the
link length goes to zero. In this limit, the energy is independent of
the link length $\ell$. But this will not be the case in general
statistical mechanics theories.  The behavior of the theory, and
experimental obeservables, depend on the fundamental length scale
$\ell$. There is, however, a

The systematic study of the length scale dependence of statistical
mechanics theories is called Renormalization Group theory \cite{Fisher1998}. If we
confine our interest to the description of the tangents of the
polymer,}
there is only one renormalizable term in the energy
density with the right symmetries to meet our assumptions. 

The
fact that more general energy densities always have continuum limits
describable by WLC models does not mean, however, that the WLC
exhausts the physically legitimate and interesting models for stiff
polymers. After all, we cannot expect continuum elasticity to remain
valid on molecular length scales.
Rather, this observation only implies that models more
general than WLC must be defined with respect to some finite length scale $\ell$.
The assumptions we made in \sref{ss:lef} imply that 
the partition function for an \pnalt{unconstrained}{} $N$ link
chain decouples into $N-1$ factors of the \pnalt{single-vertex}{}{} partition
function:
\begin{equation}
q \equiv \int d\vec{t}_{i+1}\ e^{-E_i},
\end{equation}
where $\vec{t}_{i+1}$ is the out-going tangent of link vertex $i$ (see
\fref{gdiscretechain}). In the expression above, \pnalt{$\int
d\vec{t}_{i+1}$ denotes an element of the $d-1$-dimensional sphere of
unit vectors in $d=2$ or 3 dimensions;}{we have written
the sum over the final tangent implicitly since we do not wish to
limit ourselves to a particular dimension} both two and three
dimensions are of experimental interest.

We now introduce the fundamental tangent distribution function.
The tangent distribution function is the conditional probability
density for the final tangent, given an initial tangent. The
fundamental tangent distribution function is the distribution
function over just one link, length $\ell$, and is related to the
vertex energy by the Boltzmann Distribution (\eref{gboltzmann})
\begin{equation}
\pnlabel{gftdfd}
g(\vec{t}_{i+1},\vec{t}_i) \equiv q^{-1}e^{-E_i},
\end{equation}
where $\vec{t}_{i}$ and $\vec{t}_{i+1}$ are the initial and final
tangent respectively and the deflection angle at vertex $i$ is given by
$\cos\theta_i = \vec{t}_i\cdot\vec{t}_{i+1}$. The chain statistics of the theory are completely determined by the fundamental tangent distribution.  

\subsection{Sub-Elastic Chain Model\pnalt{}{: A toy model}\pnlabel{gSECM}}
We now introduce an explicit toy model for DNA bending that
differs dramatically from the WLC model. Although the symbolic results
below can be applied to the analysis of any of the semiflexible polymer
models specified by \eref{pne1}, we will illustrate the method by using
the toy model to compute experimental observables
like  force extension,  the cyclization $J$ factor, etc.\ for an
explicit generalized theory. The model we will study has a bending
energy that is softer for high curvature than the WLC model.
Nevertheless, it reproduces the successful long-length-scale predictions of WLC.

We have already described one such model in an earlier paper
\cite{wiggins1} and a similar model was also proposed by Yan and
Marko \cite{Yan2004,YanPRE}. In both cases, the high curvature softening was
introduced by allowing kinking, or curvature localization: beyond a
critical strain, the DNA's resistance to bending was supposed to fall 
suddenly to zero, or some small value. Although these kinking
models reproduced the two desired features mentioned above, they
predicted that highly curved DNA should be generically kinked
\cite{wiggins1}. However, atomic force microscopy (AFM) imaging of small
minicircles generically shows them as round (although kinking can be
induced in unusual ionic conditions) \cite{Han1997a,Han1997b}.
Moreover, tightly looped DNA shows enhanced sensitivity to DNAse digestion
that is not concentrated on a single kink point, but rather is
spread throughout the loop \cite{hoch86a}. Finally, recent
molecular-dynamics simulations of DNA minicircles show the spontaneous 
formation of sharp bends without strand separation \cite{laveryPrivate}. For these
reasons, this paper will explore a class of models with nonlinear DNA
elasticity but without the catastrophic breakdown characteristic of
kinking.


The bending energy functions we will study
comes from the observation,
well known in continuum mechanics, that a rod bending energy density that
is non-convex in curvature induces kinking when the rod is strongly bent
\cite{rjames}. To avoid kinking, we must therefore require that our
effective bending energy density be everywhere a convex function of
curvature, at least on length scales observable via electron microscopy (EM) or AFM
imaging.

A simple choice of bending energy function that meets all the
conditions mentioned, but is radically different from the WLC model,
is:
\begin{equation}
\pnlabel{gcurvA} \epsilon(\kappa) = A |\kappa| .
\end{equation}
which defines a family of models parameterized by $A$ and $\ell$ 
that we call ``sub-elastic chain'' (SEC) models. We will show that taking
$A=5.3$ and $\ell=5\,$nm gives rise to a model with the persistence length 
$\xi=53\,$nm needed to describe the long-scale behavior of DNA in moderate-salt solution \cite{pnget}. As a final motivation, AFM studies of the tangent-tangent correlation of DNA adsorbed to mica appear to fit a bending energy of roughly this functional form \cite{VanNoort}. 

The \SEC
model illustrates several of the points we wish to make. In particular, it is  clear that high curvature, the bending stiffness in
this \SEC model is softer than the corresponding WLC model at high deflection
(we shall show that the persistence length of, and the energy is
nowhere non-convex. Nevertheless, we will see that the SEC model
reproduces essentially the same behavior as WLC for those aspects of
DNA mechanics that have been well tested. 



\pnalt{We emphasize that the \SEC{} model defined by \eref{gcurvA} is a toy
model, and not intended as a realistic, accurate representation of DNA. In particular, the nonanalytic behavior of \eref{gcurvA} at $\kappa=0$ is not meant to be taken literally. 
Instead, it illustrates our calculational method, and our larger point
that the classic DNA-mechanics tests  underdetermine even the
coarse-grained effective DNA mechanics on biologically relevant length
scales.
}{
The reader may have several quite substantive objections to the claim
that \eref{gcurvA} correctly describes the bending energy of DNA in
solution. These measurement were performed on DNA, adsorbed to mica at
low Magnesium concentrations, then rinsed and dried before the DNA
conformations were scanned using an AFM \cite{VanNoort}. It is
believed that (i) the adsorption of DNA to the mica substrate is weak
enough that the chain can equilibrate on the surface and that (ii) the
mica does not significantly affect the bending energy of the chain
because the persistence length of the polymer is roughly the same as
that measured in solution \cite{Rivetti1996}. There are also important
limitations on the accuracy of these measurements due to the AFM tip
convolution, pixillation, and tracing algorithm used to extract the
bending angles \cite{VanNoort}. Another objection to the proposed
model is that the energy is non-analytic at zero curvature.

We wish to use these experiments only to motivate a class of models in which the bending energy becomes softer than quadratic at large deflection. For the moment, let us lay our fears aside and analyze the SEC model to see what the consequences of the bending energy described by  \eref{gcurvA} are. Is this model, which is dramatically different from the WLC model on short length scales, compatible with other measurements of DNA chain statistics on long length scales as suggested by Renormalization Group arguments? Can this type of model reproduce the anomalously high $J$ factor observed by Cloutier and Widom \cite{CW}? 
To answer these questions, the detailed form of the bending energy will not be particularly important. For instance, we could introduce a small quadratic regime to the bending energy without significantly affecting the predictions of the model or the fit to experiment.  In the remainder of \sref{gsdeftheor}, we shall explicitly demonstrate that the tangent distribution function of the \SEC model approaches that predicted by the WLC model as the contour length of the chain increases.
}

\pnalt{}{\pfign{figures/genmodel/vanNoort6}{gvanNoort}{}{The bending energy of DNA on short length
scales?  The tangent distribution function is measured as a function of 
deflection angle for 5 nm sections of long sequences of DNA adsorbed 
to mica via AFM \cite{VanNoort}. The bending energy deduced from the 
tangent distribution function (\eref{genergytdf}) (black dots) is significantly non-harmonic and
can be approximately fit to the functional form $A|\theta|/\ell$ (red
curve). For comparison, the WLC bending energy is also shown
(dotted blue curve) for persistence length 53 nm which accurately
describes the long-length-scale DNA statistical mechanics but fits
the short-length-scale experimental data very poorly.  We shall
call the model based on the fit to the experimental data the \SEC
model. {\it Note to reader:} See also ???
for a fit to unpublished AFM data. }
}

\subsection{Measurements of the short-length-scale bending
energy\pnlabel{gFRET} }
\label{s:mslsbe}
The force required to tightly bend short sections of DNA has recently been directly measured by Liphardt and coworkers \cite{Shroff2005} via a fluorescence resonance energy  transfer (FRET) force sensor.   In this experiment, a sequence of DNA 9.18 nm in length is tightly bent by a linking sequence of single-stranded DNA as illustrated in \fref{liphardt}. This contour length is represented in our theory by two links ($\ell=5$ nm) and a single vertex. The deflection angle is roughly $2\pi/3$. It is straightforward to estimate the deflection force in both the discrete WLC and SEC models:    
\begin{eqnarray}
f_{\rm SEC} &\approx&  \frac{A}{\ell \cos \pi/6} \approx 5.5\ {\rm pN} \\
f_{\rm WLC} &\approx&  \frac{\xi \theta}{\ell^2 \cos \pi/6} \approx 25\ {\rm pN}.
\end{eqnarray}
(In this estimate, we have used $\ell = 4.6$ nm, half the contour length of the dsDNA.) The experimentally measured force, $6\pm 5$ pN, is approximately equal to that predicted by the \SEC model but is more than a factor of two smaller than that predicted by the elastic rod model (WLC). These experiments again indicate that the WLC model fails to describe the high-curvature bending of short sequences of DNA. At least at this deflection angle, the SEC model approximately predicts the deflection force. Note that
if the kinking model of \rrefs{Yan2004,wiggins1} literally described short sequences of DNA, this force would be {\it zero}, contrary to the 
experiment---another motivation for our introduction of generalized elasticity models.

\pfign{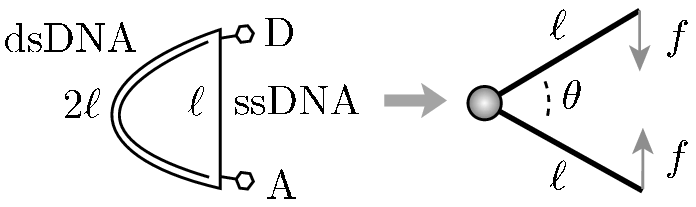}{liphardt}{}{Measuring the deflection force of highly-bent short sequences of DNA using a FRET force sensor \cite{Shroff2005}. Cyclized sequences of single-stranded DNA are hybridized with shorter complementary sequences. The single-stranded region of DNA acts as a force sensor. The external force is measured by the FRET efficiency of FRET dyes (D and A) positioned at either end of the force sensor. For a rough estimate, we model this molecule as two links under a deflection force load $f$ induced by the single-stranded DNA linker. The deflection angle $\theta$ is roughly $2\pi/3$ since the single-stranded DNA is roughly the same length as the link length.}

\subsection{The propagator and composition\pnlabel{gprop}}

The locality assumption in the definition of the bending energy implies that each vertex bends independently. The fundamental tangent distribution function is the conditional probability of a final tangent, given an initial tangent for a single vertex. Computing the tangent distribution functions for chains of several links is therefore straightforward. These conditional probabilities are simply the product of conditional probabilities for single vertices, summed over the orientations of the intermediate tangents \cite{YanPRE}
\begin{equation}
\pnlabel{glongcomp}
G(\vec{t},\vec{t}\,';N\ell) = \int \underbrace{d\vec{t}_1...d\vec{t}_{N-2}}_{N-2}\  g(\vec{t};\vec{t}_1)\, \underbrace{g(\vec{t}_1;\vec{t}_2)...g(\vec{t}_{N-3};\vec{t}_{N-2})}_{N-2} \, g(\vec{t}_{N-2};\vec{t}{\,}'),
\end{equation}
where we have written the $N$ link tangent distribution function as a function of the arc length, $N\ell$.
This notation is needlessly cumbersome. It is therefore convenient to introduce the propagation operator (or transfer matrix \cite{YanPRE})
\begin{equation}
\pnlabel{gcalG}
{\cal G} \equiv \int dtdt' \ \left|\vec{t}\,\right>
g(\vec{t},\vec{t}\,') \left<\vec{t}\,'\right|,
\end{equation}
where $\left<\right|$ and $\left|\right>$ is the canonical bra ket
notation of statistical mechanics (or quantum mechanics) \cite{sakurai}. These states are a continuum basis:
\begin{equation}
\left<\vec{t}\ |\vec{t}\,'\right> =
\delta\left[\vec{t}-\vec{t}\,'\right],
\end{equation}
where $\delta$ is the Dirac delta function on the space of unit tangent vectors.


The propagation operator, $\cal G$, applied on a state gives the
state (probability distribution) after one additional link. This property is called
composition and is a direct result of the locality discussed above. We can now 
rewrite \eref{glongcomp} more concisely
\begin{equation}
G(\vec{t};\vec{t}\,';N\ell) = \left<\vec{t}\,\right|{\cal
G}...{\cal G}\left|\vec{t}\,'\right> = \left<\vec{t}\,\right|{\cal
G}^N\left|\vec{t}\,'\right>,
\end{equation}
where the weighted sum, or
path integral, over all intermediate configurations is now implicit.  
By changing the basis in the next section, we
shall show that this expression is also a convenient computational
tool for understanding general theories \cite{YanPRE}.



\subsection{Symmetry considerations}
The tangent basis we have exploited to write the tangent
distribution function is not particularly convenient
computationally since the operator is not expressed in its eignenbasis 
in which it is diagonal.  To find an eigenbasis for this
operator, we exploit the rigid body rotational invariance of the
tangent distribution function.  In $D$ dimensions, the rigid-body-rotational 
invariance of the model
implies that the propagator commutes with the generators of rotation
\begin{equation}
\left[{\cal G},{\cal L}_{ij}\right] = 0,
\end{equation}
where ${\cal L}_{ij}=-{\cal L}_{ji}$ are the generators of rotation in the $ij$ plane. 
The propagator therefore also commutes with the Casimir operator, 
which in Quantum Mechanics would correspond to the total angular momentum:
\begin{equation}
 {\cal L}^2 \equiv {\textstyle \frac{1}{2}}\sum_{i,j=1}^D {\cal
L}_{ij}{\cal L}_{ij}.
\end{equation}
Since ${\cal L}^2$ and $\cal G$ commute, they share an eigenbasis \cite{sakurai}.
The angular momentum states span the tangent space and are eigenvalues of ${\cal L}^2$:  
\begin{equation}
{\cal L}^2\left|l\,{\bf m}\right> = l(l+D-2) \left|l\,{\bf m}\right>, 
\end{equation}
where $l$ is the total angular quantum number and we write the other 
angular quantum numbers collectively as ${\bf m}$.
The propagator can therefore be expanded in this eigenbasis \cite{YanPRE}
\begin{equation}
\pnlabel{ggl}
{\cal G} = \sum_{l{\bf m}} g_{l}\left|l{\bf m}\right>\left<l{\bf
m}\right|,
\end{equation}
where the $g_l$ are coefficients that depend only on the quantum number
$l$ but not on ${\bf m}$. 
\eref{ggl} is the desired diagonalization of the propagator $\cal G$. 




Explicitly, in two dimensions, it is convenient to use the eigenfunctions \cite{sakurai}
\begin{equation}
\left<\vec{t}\,|l m\right> = \frac{1}{\sqrt{2\pi}}\exp( -im\theta) ,
\end{equation}
for integer $m$ and $l\equiv|m|$. Note that the quantum number $m$ is sufficient to describe the state but we have introduced a second quantum number, $l$, which is invariant under a generalized notion of rotational invariance in two dimensions, including the discrete transformation $\theta\rightarrow -\theta$ (parity inversion).

In three dimensions, it is convenient to use the eigenfunctions \cite{sakurai}
\begin{equation}
\left<\vec{t}\,|lm\right> = Y_{lm}\left(\theta,\phi\right),
\end{equation}
where the $Y_{lm}$ are the spherical harmonics. In this case  $m$ is the eigenvalue of the $z$ component of the angular momentum operator ${\cal L}_{12}$.

The orthonormality of the basis implies that the $g_l$ are uniquely determined and can be
found in the usual way (\aref{a:a} \eref{ggl2d} and \eref{ggl3d}). 
It is now straightforward to perform the N+1 link path integral of \eref{glongcomp}  
\begin{equation}
{\cal G}^N = \sum_{l\bf{m}} (g_{l})^N\left|l{\bf
m}\right>\left<l{\bf m}\right|,
\end{equation}
since the propagation operator is diagonal. 
 
We return now to the SEC model proposed in \sref{gSECM}.
The $N$ link tangent distribution function for the \SEC model is shown in 
\fref{gEvolVanNoort}. This figure explicitly illustrates the scale dependence 
of statistical mechanics theories. For short-contour-length chains, the WLC and \SEC
theories make dramatically different predictions, but as the contour length of 
the chain increases, the differences between the distribution functions of the two theories decrease until at long contour length, the theories are essentially indistinguishable. This is the essence 
of the renormalization group: at short length scales, the mechanics of the chain can be extremely complicated but the thermal fluctuations sum over many intermediate configurations and hide the underlying complexity on longer length scales. We shall show this for general theories in \sref{gspersistence}.

\pfign{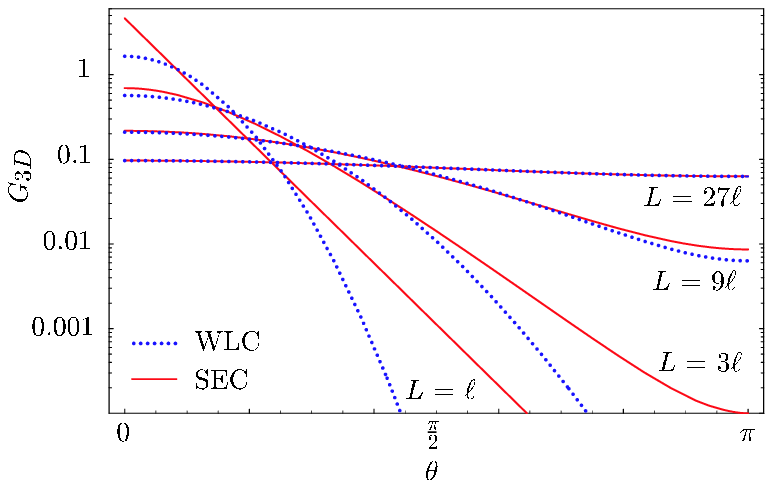}{gEvolVanNoort}{}{ Evolution of the 3-dimensional tangent distribution function $G_{3D}(\vec{t};\vec{t}{\ }';L)$ with increasing separation $L$.
In the figure above, the WLC and \SEC tangent distribution functions are plotted 
as a function of the deflection angle $\theta$ for several contour lengths.
The linear dependence of \SEC bending energy on the deflection angle, visible in the fundamental distribution
function ($L=\ell=5$ nm), is lost at longer contour length. For
$L\gg\ell$, the tangent distribution approaches the WLC
distribution function with a persistence length of 53 nm despite
dramatically different behavior at short contour length. This
loss of the short length structure of the tangent distribution
function is universal and explains the success of the WLC model in
describing many semiflexible polymer phenomena. }

\subsection{Contour length continuation}
Since we will frequently be interested in the properties of the
polymer on length scales much longer than the fundamental link
length $\ell$, it is useful to introduce a Hamiltonian operator
defined by
\begin{equation}
\pnlabel{gHOe}
{\cal H} \equiv - \ell^{-1} \log {\cal G} = \sum_{l{\bf m}}
h_{l}\left|l{\bf m}\right>\left<l{\bf m}\right|.
\end{equation}
{\cal H} is also diagonal in the angular momentum representation with eigenvalues
$h_l= -\ell^{-1} \log g_l$. 
We
call this operator the Hamiltonian operator because in the WLC
model, the statistical mechanics of the chain corresponds to a
quantum particle on a $D-1$ sphere. The tangent distribution
function is equal to the quantum propagator where time has been
continued to imaginary arc length. The operator ${\cal H}$ is
equal to the Hamiltonian of the corresponding quantum mechanical particle system.

We can rewrite the $N$ link as
propagator
\begin{equation}
{\cal G}^N = \exp( -{\cal H} N\ell ),
\end{equation}
where $N\ell$ is the contour length corresponding to $N$ links.
The advantage of this reformulation of the distribution function is
that it introduces a natural extension to fractional numbers of
links by replacing $N\ell$ by the contour length $L$ defined for
all positive real numbers: 
\begin{equation}
\pnlabel{gpropop} {\cal G}\left(L\right) \equiv \exp( -{\cal H} L),
\end{equation}
although rigorously, it is understood that this function is only
defined for contour lengths equal to an integral number of links.

\subsection{The meaning of persistence length, and the stiff-polymer limit
\pnlabel{gspersistence}}
What is the meaning of persistence length in general models like the ones we have described? Persistence
length describes the length scale on which the polymer maintains
its tangent orientation.  For the WLC model in $D$
dimensions, the tangent persistence is
\begin{equation}
\pnlabel{gpersistencelength} \left<\vec{t}(0)\cdot\vec{t}(L)\right> =
\exp\left[ -L(D-1)/2\xi\right],
\end{equation}
where $\xi$ is the persistence length, which also appears in the energy
density as the bending modulus in \eref{gWLCED}. In general models, the tangent persistence
(\eref{gpersistencelength}) 
has the same functional form but $\xi$ no longer corresponds to a 
bending modulus. 
We shall therefore simply use \eref{gpersistencelength} to define the persistence length of general models.

The tangent persistence corresponds 
to the $l=1$ mode of the propagator.
(In the quantum mechanical correspondence, $\vec{t}$ is a vector and creates a state of spin one.)
Comparing 
\erefs{gpropop} and \ref{gpersistencelength}, the persistence length
is
\begin{equation}
\pnlabel{gpl2}
\xi_D \equiv (D-1)/2h_1,
\end{equation}
where $h_1$ is the $l=1$ eigenvalue of the Hamiltonian. 
Note that we have explicitly written a subscript $D$ to denote the dependence on 
dimension. In the WLC model, $\xi$ is independent of dimension, 
but in more general models this is not the case. 


The persistence length also controls the long-length
characteristics of the polymer. The mean-squared end-to-end
distance can be written in terms of the tangent persistence
\begin{equation}
\left<\vec{X}^2\right> = \int_0^Ldsds'\
\left<\vec{t}(s)\cdot\vec{t}(s')\right>.
\end{equation}
Since \eref{gpersistencelength} applies to both the WLC model and general models, the dependence of the mean squared end-to-end distance on persistence length and contour length is identical to the WLC model. The
same is true for radius of gyration, which can also be written in terms of an integration of \eref{gpersistencelength}.  It is also well known that
the long-contour-length spatial distribution of semiflexible polymers is described by
the Gaussian Chain model \cite{doi1986}.  The width of the Gaussian distribution is
determined by the mean squared end-to-end distance; the
relation between the Kuhn length and the persistence
length is therefore the same for our  general models as for the WLC model.

We can immediately exploit \eref{gpl2} and \eref{ggl3d} to analyze the \SEC model. 
The persistence length, computed for the \SEC 
model in three dimensions is 53 nm which matches solution measurements.

\noindent{\sl Stiff-polymer limit: }
We now examine the tangent distribution function in the stiff polymer
limit and show that the WLC model is universal at long contour
length as predicted by the Renormalization Group.  Our explicit 
computations of the \SEC tangent distribution function in \sref{gprop} have already provided 
one explicit example of this behavior, but we address this question generally in this section.
%

By definition the stiff polymer limit implies that the fundamental tangent distribution function, $g$,
is narrowly distributed around zero deflection. We will exploit this fact by expanding the basis
functions in the deflection angle and computing the eigenvalues of the propagator (\eref{ggl}) 
to lowest order in the deflection angle. In dimension
$D$, this calculation, although straightforward, requires some technical mathematics. 
We therefore relegate the details of this calculation to the appendix, \sref{gstiffpol}, and present only the results.

The propagator in the stiff polymer limit
is \eref{sple}
\begin{equation} 
{\cal G} = 1-\frac{\ell}{2\xi}{\cal L}^2 +{\cal O}[{\cal
L}^4(\ell/\xi)^2],
\end{equation}
where $\xi$ is the persistence length defined by the $l=1$ eigenvalue of the Hamiltonian operator
(\eref{gpl2}). Note that the ${\cal L}^2$ term is understood to be small for small values of the angular quantum number $l$ since, in the stiff polymer limit, the link length $\ell$ is much shorter that the persistence length $\xi$.
The corrections are order ${\cal L}^4\left<\theta^{4}\right>$ and scale as $l^4$ for large $l$. Clearly this approximation holds only
for small angular quantum number $l$. It is convenient to compute 
the Hamiltonian operator 
\begin{equation}
\pnlabel{grenorm12}
{\cal H} =\frac{1}{2\xi}{\cal L}^2 +{\cal O}[{\cal
L}^4(\ell/\xi)^2],
\end{equation}
which is identical to the WLC Hamiltonian to lowest order in the deflection angle. Again, 
the correction scales like $l^4$ which implies that this relation holds only for small angular quantum numbers. 

The correspondence between the Hamiltonian operators for general models and the WLC model for small angular quantum numbers implies that the long-contour-length behavior of the polymer is universal and determined by the persistence length alone. This correspondence is shown explicitly for the \SEC and WLC theories in \fref{gHlfig}. At long contour length, only states with small $l$ contribute since higher-$l$ contributions decay quickly.  Remember that the propagator is
\begin{equation}
{\cal G}  = \exp( -{\cal H}L),
\end{equation}
and the eigenvalues of ${\cal H}_{\rm WLC}$ scale as $l^2$ for large $l$. 
The tangent distribution function is therefore well approximated by the WLC model at long contour length: 
\begin{equation}
\lim_{L\gg\ell} {\cal G}(L) = {\cal G}_{\rm WLC}(L).
\end{equation}
The details of the short-length-scale bending energy affect only the large $l$ eigenvalues of the Hamiltonian operator and are therefore irrelevant at long contour length, as predicted by the renormalization group. 

Although we have yet to compute the spatial distribution function, we have explicitly shown that measurements that are only sensitive to the long-length-scale chain statistics do not determine the short-length-scale behavior of the theory and that violations of the wormlike chain model, while disguised by thermal fluctuations at long contour length, are generic as the length scales probed by experiment approach the fundamental or structural length scale of the chain. 

\pfign{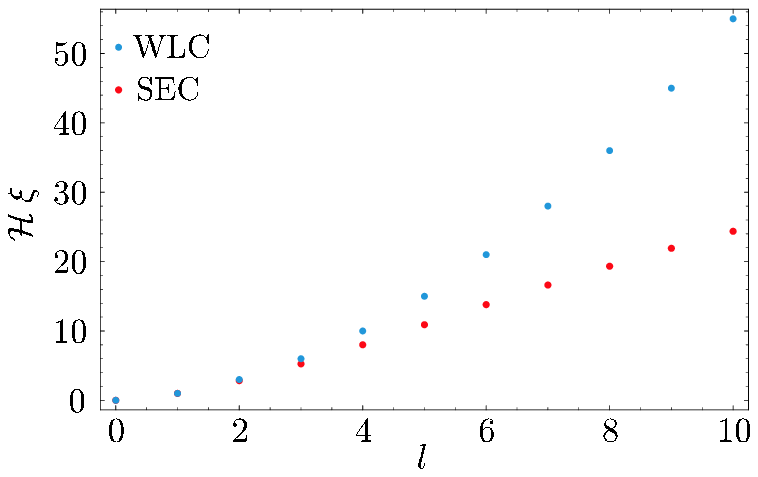}{gHlfig}{The eigen-spectrum of ${\cal H}$ for the SEC and WLC models}{The eigen-spectrum of ${\cal H}$ for the SEC and WLC models. 
The eigenvalues of the Hamiltonian operator for the WLC and SEC theories are compared 
 as a function of the angular quantum number $l$. Both theories have an identical persistence length, $\xi=h_1^{-1}=53$ nm.  The eigenvalues of the Hamiltonian are coincident for small $l$ but diverge as $l$ increases. The $l$th moment of the distribution function decays as $\exp(-h_lL)$. The larger eigenvalues of the Hamiltonian, for which the two theories differ, are therefore relevant only for small $L$, implying that the SEC and WLC chain statistics are identical for long-contour-length chains.}

%



\section{The spatial distribution}
\pnlabel{gssdf}
For most applications, it is the spatial distribution of the polymer rather than the tangent distribution function which is of phenomenological interest. From solution scattering to force-extension, cyclization to looping, the spatial distribution function is directly observable. In this section, we shall develop a near exact formalism for computing the spatial distribution function. Our focus will be exclusively three dimensions but computations in other dimensions are a simple extension of the methods discussed here. 

The tangent-spatial distribution function is the probability density of  end displacement $\vec{X}$ and final tangent $\vec{t}_f$ given an initial tangent  $\vec{t}_i$ for an arc length $L$ chain. It is convenient to write the tangent spatial distribution in terms of the spatial delta function \cite{YamakawaBook}
\begin{equation}
\pnlabel{gsp}
G(\vec{X};\vec{t}_f,\vec{t}_i;L) = \left<\vec{t}_f\right|\exp\left[-{\cal H} L\right]\delta^3[\vec{X}-\int_0^Lds\  \vec{t}(s)]\left|\vec{t}_i\right>,
\end{equation} 
where we have written the distribution function in the continuum limit.
We shall reintroduce an operational definition of this continuum limit in a moment. 

To compute the tangent-spatial distribution function, we introduce an operator-valued spatial distribution function \cite{Andy2}:
\begin{equation}
\pnlabel{gopvalsdf}
{\cal G} (\vec{X};L)= \int d\vec{t}d\vec{t}\,'\ \left|\vec{t}\ \right>G(\vec{X};\vec{t},\vec{t}\,';L)\left<\vec{t}\ '\right|,
\end{equation}  
which allows us to keep the tangents implicit in our expressions. We shall call this operator the spatial propagator since it obeys the composition property of Green Functions:
\begin{equation}
\pnlabel{gconvol}
{\cal G} (\vec{X};L+L') = {\cal G} (\vec{X};L) \otimes {\cal G} (\vec{X};L'),  
\end{equation}
where $\otimes$ denotes the spatial convolution.

As before, it will be convenient to work in the angular momentum basis with the matrix elements
\begin{equation}
\pnlabel{gmatrixelement}
{G}_{lml'm'}(\vec{X};L) \equiv \left<l\ m\right| {\cal G}(\vec{X};L) \left| l'\ m'\right>,
\end{equation}
since this basis diagonalizes the Hamiltonian (although not the spatial propagator). 
Finding the spatial propagator reduces to the ability to explicitly compute all the ${G}_{lml'm'}$. 

We shall be able to derive exact expressions for the Fourier-Laplace Transform of the spatial propagator in the continuum theory in terms of the transformed matrix elements (\eref{gmatrixelement}). We adopt the Fourier Transform convention
\begin{equation}
G(\vec{k};L) \equiv \underset{X\rightarrow k}{\cal F}\ G(\vec{X};L) \equiv \int d^3 X\ G(\vec{X};L)\ \exp( -i\vec{k}\cdot\vec{X}),
\end{equation}
and the Laplace transform convention
\begin{equation}
\tilde{G}(\vec{k};p) \equiv \underset{L\rightarrow p}{\cal L}\ G(\vec{k};L) \equiv \int_0^\infty dL\ G(\vec{k};L)\ \exp( -pL).
\end{equation}
The derivation of the transformed matrix elements exploits the same techniques used recently by Spakowitz and Wang \cite{Andy,Andy2,Andy3} to derive exact results for the WLC model. The extension of these results to the generalized theories considered here is straightforward. We shall therefore include only a brief derivation in \aref{gtsp} although we discuss the results in the main text.

It is important to note at this stage that the results derived for the spatial distribtution, although derived in a method analogous to that exploited in \rref{Andy2}, will not be exact solutions to generalized discrete-link models. Rather, the results quoted here are exact-solutions to the analytically continued theories defined by \eref{gpropop}. 
That is, we have assumed a formulation of the discrete-link theories that defines the tangent distribution function for all $L>0$, although formally this distribution function is defined only for contour lengths equal to an integral number of links.  For semiflexible polymers longer than a few links, this is an excellent approximation. (For instance, the discrete and continued theories are later compared in \fref{gk}.) We have therefore called this solution ``near-exact'' in the text. 

\subsection{The spatial distribution function}

\pnlabel{gtsdfs}
In force-extension and solution scattering experiments the tangents of the polymer are not directly probed by experiment; it is only the spatial distribution function rather than the tangent-spatial distribution function which is observed. We shall therefore introduce the spatial distribution function, $K(\vec{X};L)$, which is defined as the probability density that a contour length $L$ polymer has end displacement $\vec{X}$. The spatial distribution function is the tangent-spatial distribution function summed over the final tangent and averaged over the initial tangent:
\begin{equation}
K(\vec{X};L) \equiv \frac{1}{4\pi} \int d\vec{t}_f d\vec{t}_i\ G(\vec{X};\vec{t}_f,\vec{t}_i;L) = G_{0000}(\vec{X};L),
\end{equation}
where the last equality expresses the spatial distribution function in terms of a matrix element of the spatial propagator. 

The Fourier-Laplace transform of this matrix element, a continued fraction, is computed in \eref{gpropmatrixelement}. The explicit expression for the transformed spatial distribution function is
\begin{eqnarray}
\pnlabel{gkprop}
\tilde K(\vec{k};p) &=& \frac{{\displaystyle1}}{{
\displaystyle p+h_0+\frac{B_1 k^2}{
p+h_1+\frac{B_2k^2}{\cdots}
}
}},
\end{eqnarray}
where the $h_l$ are the eigenvalues of the Hamiltonian operator, \eref{gHOe}, and the $B_n$ coefficients are defined as 
\begin{equation}
B_n  \equiv \frac{n ^2}{4n ^2-1}.
\end{equation}
This expression is identical to that derived for the WLC model \cite{Andy}, except that the eigenvalues of the Hamiltonian operator, $h_l$,  are those for a generic theory rather than the WLC eigenvalues. 
Otherwise the expression is unchanged. 

\pfign{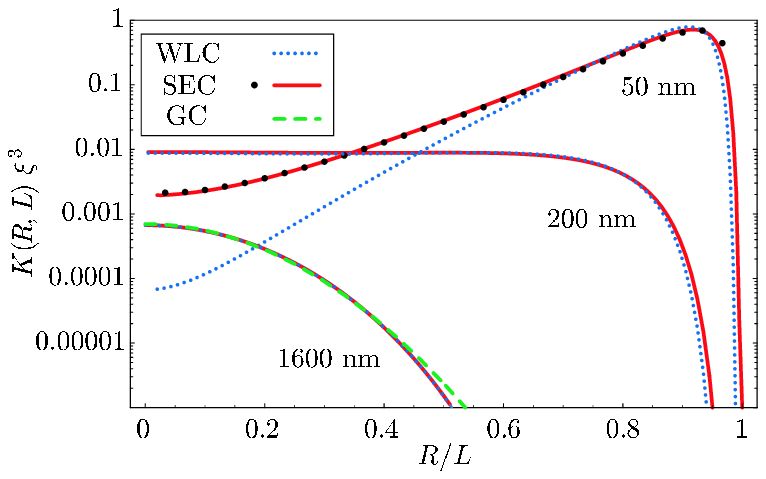}{gk}{The spatial distribution (distribution of spatial separation $R$) for the WLC and \SEC theories}{The spatial distribution for the WLC and \SEC theories. All curves except the black dotted curve have been computed using the inverse transform technique.  To check the validity of the validity of the technique, the black dots show a direct Monte Carlo integration for the shortest contour length \SEC curve (red). We have chosen the contour lengths of the chains to illustrate two types of renormalization. At 50 nm for large deflection ($R/L\sim 0$), the \SEC (solid) and WLC (dotted) theories differ by two orders of magnitude. For a 200 nm contour length, \SEC and WLC predict nearly identical distributions, but this distribution is clearly not Gaussian. For long contour length, however these theories renormalize to the Gaussian chain model (dashed).   }

The spatial distribution functions for the WLC and SEC models are plotted in \fref{gk}
for several contour lengths. The numerical techniques applied in this computation are described in \aref{numericalsd}.
These results again display the renormalization 
group flow toward the WLC model at long contour length. Although the two theories make dramatically different predictions for short-contour-length chains, the predictions coincide at long contour length! 


The suppression of the short-length-scale structure of the theory can again be understood in terms of the eigenvalues of the Hamiltonian operator. The levels of the continued fraction (\eref{gkprop})
can be understood as contributions from transitions to states of increasing angular quantum number $l$. But these high-angular-momentum states decay quickly due to their large eigenvalues of the Hamiltonian. 
We can also understand the irrelevance of high-angular-momentum states at long length in terms of the wave number $k$. Long length scales correspond to small wave number. The levels of the continued fraction are multiplied by $k^2$ and are therefore suppressed for small wave number, implying that the higher-angular-momentum states have successively less relevance at long length scales.


It is also instructive to consider the long-length-scale limit of the spatial distribution function
since we know that this limit is described by the gaussian chain model. The long-length-scale limit corresponds to the limit of small $k$ and contour dual number $p$. In this limit, the transformed spatial distribution function is
\begin{equation}
\pnlabel{gkprop2} \tilde K(\vec{k};p) \rightarrow \frac{1}{p+A_1
\vec{k}^2/h_1},
\end{equation}
which is just a Gaussian distribution with a Kuhn length of twice
the persistence length (\eref{gpersistencelength}) as we have
already argued from computations of the mean-squared end-to-end distance and has also been shown schematically for the SEC model in \fref{gk}.  



\subsection{Force-extension}
\pnlabel{gfexts} 
The force-extension of single polymer molecules has long been 
the subject of experimental interest \cite{Bustamante2000,philbook}.
The experimental observable in these experiments, the extension of the polymer under an external force, can be directly computed from the spatial distribution function. 
Typically this force is applied to a bead, tethered to the polymer, using an
optical or magnetic trap \cite{Bustamante1994,stri98b,Bustamante2000}. The
restoring force against extension is entropic in nature (for inextensible polymers). 
This entropic force is induced by the reduction in the
number of micro configurations available to the chain as the extension
is increased. 

The successful comparison of WLC to single-molecule force-extension data has 
been described as the strictest
test of the WLC model \cite{Bustamante2000}. But how do other semiflexible polymer models compare? Can these models also reproduce the precise fit to experiment? To answer these questions, we next compute the force-extension for general models and explicitly compare the extension in the SEC and WLC models (\fref{gfext}).

\pfign{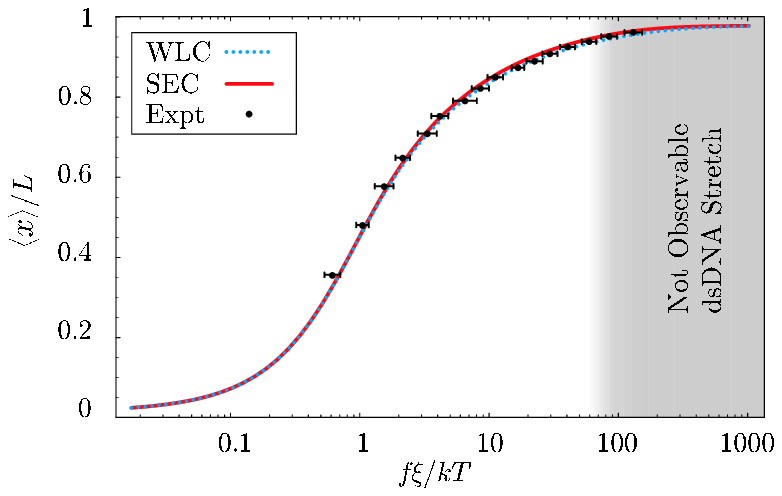}{gfext}{Force-extension}{Force-extension for the WLC and \SEC models
compared with experimental measurements \cite{pnget}. The WLC
model was fit to the experimental data to determine the contour
length and persistence length ($\xi = 53$ nm). Despite the
dissimilar short-length-scale tangent distribution function, the
behavior of the polymer under an external force is nearly
identical.  For forces greater than 10 pN, intrinsic stretching
stretch becomes important, obscuring the entropic part of the response. }

The partition function for a polymer under a constant external tension
is related to the Fourier transform of the spatial
distribution function via an analytic continuation of the wave
number:
\begin{eqnarray}
Z(\vec{f}\,) &=& \int d^Dx\
K(\vec{x};L) \ \exp\left[\vec{f}\cdot\vec{x}\right] \\
\pnlabel{gpfef} &=& \underset{L\rightarrow p}{\cal L}^{-1}\ \tilde
K(i\vec{f};p),
\end{eqnarray}
where $\vec{f}$ is the external force or tension.  The extension or mean end distance
is computed in the usual way:
\begin{equation}
\left<x\right> = \frac{\partial \log Z}{\partial f}.
\end{equation}
The force-extension for the \SEC and WLC models are compared in \fref{gfext}. 
The numerical technique applied in this computation is described in \aref{numericalsd}.

Despite the drastically different bending energy of the SEC model on short
length scales, thermal fluctuations
disguise these differences and give rise to an
extension almost identical to the WLC model. In retrospect, 
these results are hardly surprising. The theories are identical at small 
extension due to the renormalization group and at large extension due 
to inextensibility. Although, in principle, the high force limit is mathematically
equivalent to probing short length scales---they are related by analytic continuation---these 
differences are not large enough to be experimentally observable.  Physically, the rare 
high curvature bending regime, where the difference between the models is most pronounced,
is further suppressed by the application of tension.   For the study of DNA mechanics, force-extension measurements do probe the persistence length and the inextensibility of DNA, but these experiments do not effectively probe DNA elasticity on the length scales of interest for many biological processes. 
%



\subsection{Structure factor}
\pnlabel{gssf}

Another experimental observable used to characterize polymers is
the structure factor, measured by static light scattering,
small-angle X-Ray scattering, and neutron scattering experiments.
Measurements of the structure factor can probe the polymer
configuration on a wide range of length scales. Symbolically the
structure factor is
\begin{equation}
S(\vec{k}) \equiv \frac{1}{L^2} \int_0^Ldsds'\
\left<e^{i\vec{k}\cdot\left[\vec{X}(s)-\vec{X}(s')\right]}\right>,
\end{equation}
where $\vec{X}(s)$ is the position of the polymer at arc length
$s$ and we have included an extra factor of the contour
length in the denominator to make the structure factor
dimensionless \cite{Andy}. At high wave number, the structure
factor is sensitive to short-length-scale physics, whereas the
contour length and radius of gyration are probed by low wave
number. The structure factor can be rewritten in terms of the
transformed spatial distribution function
\begin{equation}
S(\vec{k}) = \frac{2}{L^2}\underset{L\rightarrow p}{\cal L}^{-1}\left[\frac{\tilde
K(\vec{k};p)}{p^2}\right],
\end{equation}
where ${\cal L}^{-1}$ is the inverse Laplace transform which can
be computed numerically.  As mentioned above, the leading-order
contributions at small wave number are determined by the polymer length and the
radius of gyration
\begin{equation}
LS(k) = L(1+{\textstyle \frac{1}{3}}\vec{k}^2R_g^2+...),
\end{equation}
where we have temporarily restored the length dimension of $S$. At
large $k$, both WLC and \SEC are straight, which gives
an asymptotic limit for large wave number
\begin{equation}
S(k) \rightarrow \frac{\pi}{Lk}.
\end{equation}
The structure factor is 
compared for the SEC and WLC models in \fref{gg.struct.renorm}. The numerical technique applied in this computation is described in \aref{numericalsd}.

\pfign{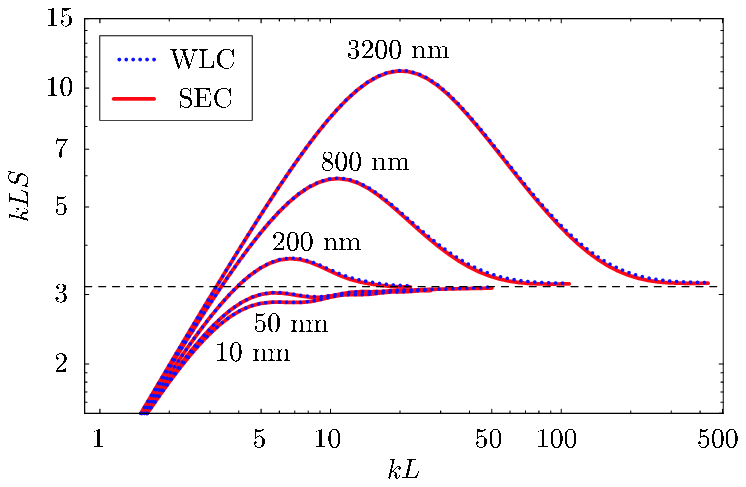}{gg.struct.renorm}{The structure factor for the SEC and WLC models}{The structure factor for the SEC and WLC models.
In the figure above, the structure factor $S$, scaled by the wave number, is plotted 
for several contour lengths. The curves are nearly identical for the two theories since the structure factor is dominated by thermally accessible configurations. Although rare, high-curvature configurations are orders of magnitude more probable in the SEC than in the WLC theory, these configurations are still too rare to significantly affect the structure factor. }

Again we find that the two theories make nearly identical predictions.
The reasoning is again similar to that explained for force-extension. The two theories 
make dramatically different predictions for rare, highly-bent configurations but the 
structure  factor is dominated not by these rare high curvature configurations but by typical thermal bending. We therefore find that the structure factor, like force extension, does not effectively probe 
the high-curvature statistics of the polymer.

%

%

\subsection{Cyclization\pnlabel{gcycls}}

The biochemical process of DNA cyclization is not in itself a
process of particular biological importance\footnote{Bacteriophages are know to cyclize their genomes after
ejection into a cell, but these genomes are typically many
thousands of base pairs and the barrier to cyclization is purely
entropic.} but cyclization experiments do provide a controlled,
bulk experimental method for probing the probability of rare,
highly-curved DNA configurations \cite{CW,Cloutier2005,Du:2005ir}.
In these experiments, linear double-stranded sequences with
complementary single-stranded ends are ligated into cyclized
sequences \cite{JacobsonStockmayer,Shore81,Shore83a,SY,Hagerman:1988hr}.
The cyclization reaction precedes via the capture of rare,
thermally activated configurations and is thought to be very similar
to the process by which looped DNA-protein complexes are formed.
Cyclization does have a very clear advantage over protein-induced DNA looping as a method of probing the high-curvature mechanics of DNA: the chain boundary conditions for cyclization (tangents aligned) are well known, in marked contrast to most DNA-protein complexes where the relevant chain boundary conditions must be determined.  

The cyclization assay is performed 
under conditions such that the  ligation reaction samples the equilibrium 
populations of unligated cyclized and oligomerized polymers \cite{Shore81}. 
The ratio of the cyclization equilibrium
constant ($K_C$) to the dimerization equilibrium constant ($K_D$)
is called the Jacobson-Stockmayer factor \cite{JacobsonStockmayer}
or $J$ factor and is proportional to the tangent-spatial
distribution function of the polymer \cite{Shore81,SY}
\begin{equation}
\label{jfactoreq}
J \equiv K_C/K_D = 4\pi G(0;\vec{t},\vec{t};L) = {\rm tr}\  {\cal G}(0;L),
\end{equation}
where $G$ is the tangent-spatial distribution function for end-to-end displacement $0$ and aligned end tangents, for a contour length $L$ polymer. The $J$ factor can also be written as the trace  of the spatial propagator. (The matrix elements of the spatial propagator are written explicitly in \aref{gtsp}.)
Physically, the $J$ factor is proportional to the
concentration of one end at the other with the correct (aligned)
orientation for hybridization.

Our analysis neglects the condition that DNA twist must also
be aligned, which requires the use of models including the twist degree of freedom. 
This additional constraint
modulates the $J$ factor with a 10.5 bp period equal to the
helical repeat. Our interest here is in the value of the $J$
factor averaged over a helical repeat for which the effects of
twist can be roughly ignored \cite{YamakawaBook}.

\pfign{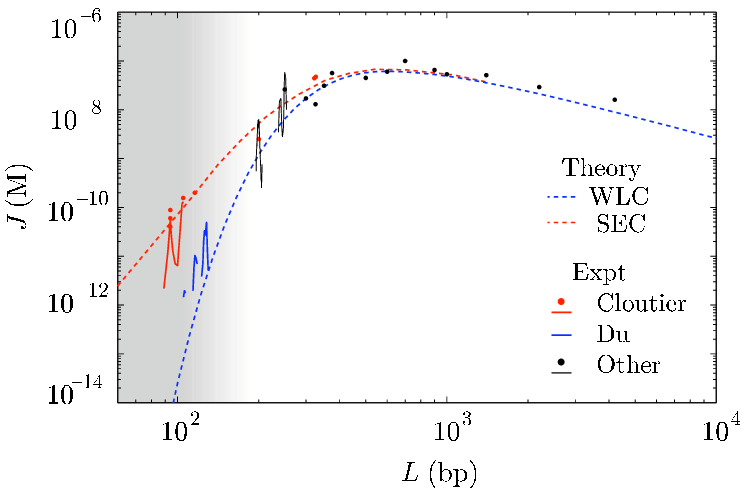}{gcycl}{%
The cyclization $J$ factor: Probing the high-curvature chain
statistics}{%
The cyclization $J$ factor: probing the high-curvature chain
statistics. In the figure above, the cyclization $J$ factor in units
of molarity is plotted for the WLC \textit{(dashed blue curve)} and SEC \textit{(dashed red curve)} models and
compared with experimental measurements \textit{(circles)}
\cite{Shore81,Shore83a,Vologodskaia2002,CW,Cloutier2005}. The
theoretical curves do not include the twist induced modulation visible
in the continuous sets of experimental data \textit{(solid curves)}
\cite{Shore83a,YamakawaBook,Vologodskaia2002,Cloutier2005}. The
renormalization group predicts that the \SEC model will be identical
to WLC for long-contour-length sequences. But, for sequences shorter
than two persistence lengths ($\lesssim$200 bp), the
short-contour-length chain statistics become important and the \SEC
$J$ factor diverges from the WLC prediction. In fact, for 94 bp
sequences, the SEC $J$ factor is three orders of magnitude larger than
that predicted by the WLC model, roughly matching the $J$ factors
measured by Cloutier and Widom \cite{CW,Cloutier2005} \textit{(red circles and
solid curves)} whereas subsequent measurements by Du {\it et al.}~\textit{(blue
solid curves)} are commensurate with the predictions of the WLC model
\textit{(dashed blue curves)}. Our results predict that a short-contour-length
anomaly in the $J$ factor is generic for sufficiently short sequences,
but the contour lengths at which the WLC model breaks down is model
dependent.}
%

\fref{gcycl} compares the cyclization $J$ factor for the \SEC and WLC theories. 
The numerical techniques applied in this computation are described in \aref{numericalsd}. The $J$ factors for sequences with contour lengths greater than two persistence lengths have long been known to match the predictions of the WLC model \cite{Shore81,Shore83a}.
For sequences shorter than two persistence lengths, the figure illustrates the short-contour-length break down of the WLC model describing the chain statistics of the SEC model.
For example, for contour lengths of roughly $0.6$ persistence lengths, which correspond to loops with
approximately the same radius of curvature as DNA bound to
histones in nucleosome complexes, the \SEC model $J$ factor is three orders of magnitude larger than predicted by the WLC model, in rough agreement with cyclization measurements of Cloutier and Widom \cite{CW}, as illustrated in \fref{gcycl}.

The qualitative picture illustrated in the \fref{gcycl} (the WLC model describes long-contour-length chain statistics, but fails at sufficiently short contour length) is the generic result from $J$ factor computations in general models. These results were qualitatively predicted by the renormalization group ideas we have discussed throughout the paper.  From an experimental perspective, the cyclization assay is clearly a powerful technique for probing the short-contour-length chain statistics of DNA. In particular, this technique has very clear advantages over force-extension and solution-scattering experiments since {\it (i)} cyclization assays probe the chain statistics of DNA in a way that is qualitatively similar to biological DNA looping applications and {\it (ii)} cyclization experiments are extremely sensitive to the differences  between models at short contour length. 


%


\subsection{Beyond the $J$ factor}
The $J$ factor is not the only effective concentration of interest. DNA looping is integral to the function of many gene regulatory proteins.  The affinity of these proteins for DNA, and therefore their function, depends sensitively on the looping free energy, or equivalently the effective concentration of the looped DNA. (For instance, see \rrefs{Muller1}, \cite{Bintu2005a}, and \cite{Bintu2005b}.) Once the geometry of the loop is known---the displacement of the binding sites ($\vec{X}$) and the orientation of the bound DNA ($\vec{t}$ and $\vec{t}\,'$)---both the \SEC and WLC models make predictions for the effective concentration:
\begin{equation}
[{\rm effective\ concentration}] =  4\pi G(\vec{X};\vec{t},\vec{t}\,';L).
\end{equation}  
These statistical mechanics predictions can then be directly compared with quantitative measurements of gene expression \cite{Bintu2005a,Bintu2005b} and {\it in vitro} experiments \cite{Finzi1995}.

More general cyclization measurements may also be performed. For instance, cyclizing sequences with two short single-stranded gaps could be used to probe short-length-scale DNA mechanics. The short single-stranded sequences are very flexible and can be approximated as free hinges.  This technique could be exploited to directly measure the spatial distribution function shown in \fref{gk} for very short sequences of DNA.

\section{Discussion\pnlabel{s:dis}}
\subsection{The SEC}
\srefs{gSECM}--\ref{s:mslsbe} introduced the SEC as a toy model for DNA bending,
motivated by several physical measurements on DNA. We proceeded to show that this
simple model exhibited the long-length-scale chain statistics of the WLC model, despite dramatically increasing the predicted probability of high-curvature configurations. In particular, we
showed that the SEC model yields a cyclization $J$
factor in agreement with the measurements of Cloutier and Widom \cite{CW}. More generally, we
argued that this type deviation from WLC behavior is generic in semiflexible chain models.


These putative deviations of DNA chain statistics from the wormlike
chain model at short contour length are quite relevant for structural 
biology, where the typical radius of curvature induced by DNA
binding proteins is on order nanometers or tens of nanometers, not
persistence lengths. For example, the radius of curvature of DNA
bound in a nucleosome complex is roughly 6 nm. The structure of this
complex shows sharp bends, but no sign of melting, consistent with our
SEC model \cite{Richmond2003}. 
Similarly, DNA looped by a gene regulatory protein is typically
bent on short length scales \cite{Muller1}. If DNA is described by the \SEC
model, these tightly-bent DNA-protein complexes are orders of
magnitude more stable than predicted by the WLC model. A quantitative
understanding of biological DNA bending therefore awaits a consistent
model of short-length-scale DNA bending.

Unfortunately, precise quantitative tests of short-length-scale DNA bending are still in
the future. Vologodskii and coworkers
recently made measurements questioning the results of
Cloutier and Widom \cite{Du:2005ir}. Their measurements
suggest that the $J$ factor agrees with that predicted by the WLC model, at
least down to a contour length of 100 bp. Widom and coworker then
repeated their own measurements, however, and have confirmed their
previous results \cite{wpriv}. Also, \sref{gFRET} mentioned that Shroff {\it et al.}~\cite{Shroff2005} also
found that linear elasticity fails at high curvature. 
At the moment, it is difficult to reconcile all these
conflicting experiments.  Instead this paper has shown that
existing experiments do not uniquely confirm the WLC;  we
have examined some of the options for theories compatible with those
experiments that appear to be understood.

\pnalt{}{asked,
are there compelling theoretical arguments for the success or failure of the WLC model for describing the cyclization of 100 bp sequences?
At what length scale do we expect the WLC model to begin to capture the statistics of DNA quantitatively?


The length scale at which the WLC model breaks down is determined by the underlying, molecular structure of DNA. In our description of generalized theories, this length scale only enters as the shortest allowed value for the link length parameter $\ell$.  Until now, we have treated the link length as a parameter of the model without discussing its physical significance. In order to understand the role of the link length, it is useful to return to the equation describing the tangent distribution of the polymer (\eref{gpropop})
\begin{equation}
{\cal G}\left(L\right) \equiv \exp (-{\cal H} L).
\end{equation}
This equation gives a recipe for computing the tangent distribution function for any contour length, based on a fundamental tangent distribution function of a contour length $\ell$ chain. But, this equation can also be evaluated for contour lengths shorter than the link length, violating the spirit of the renormalization group. These distribution functions, computed for contour lengths shorter than the link length, are not physically meaningful. On shorter length scales, the polymer will not even be described by a theory which meets the assumptions we made initially: ({\it i}) isotropic, ({\it ii}) homogeneous, and ({\it iii}) local. In fact, \eref{grenorm12} implies that if the semiflexible polymer were described by a theory satisfying ({\it i})--({\it iii}) at arbitrarily short length scales and the bending energy was everywhere finite, the chain statistics would be described by the WLC theory!
It is precisely because at least one of these assumptions breaks down on short length scales that polymers are not exactly described by the WLC model.
The generalized models discussed in this paper can describe the chain statistics before the WLC model becomes applicable.


Is the SEC model reasonable? The SEC model essentially implies
that the fundamental length scale is on order 5 nm. (If it were
significantly smaller, the chain statistics at the 5 nm length scale
would be described by the WLC model.) Such a proposal is not far
fetched since our assumption of homogeneity may fail. The link length,
5 nm, is comparable to the helical repeat of DNA. It has been
proposed, for instance, that DNA may bend preferentially into the
minor groove, an anisotropy that repeats every helical repeat
\cite{widomnuclrev,Richmond2003}. On the other hand, sequence induced
heterogeneity is probably fairly well averaged over the 5 nm length
scale since this corresponds to nearly 15 bp. Failure of the locality
assumption ({\it iii}), the break down of the nearest neighbor
interaction on these length scales, seems unlikely since it is
difficult to understand such a phenomena without bending-induced,
long-range structural changes. Such structures have not been observed
in the crystal structures of highly bent DNA on the 5nm length scale
(but see \rref{schur97a}). Based on these theoretical and structural
arguments alone, failure of the WLC model on these length scales is
neither imperative nor ruled out. Resolving  the underlying
mechanics of DNA awaits new experiments to resolve the differences
between Refs.~\cite{CW,Du:2005ir}.
}

We have repeatedly emphasized that the SEC is more a proof of
principle than a finished theory. It is a generalization of the WLC
that is extremely compact to state and can be solved almost analytically.
It shows that the classic successes of WLC can be reconciled with more
recent indications of elastic breakdown. It encodes locality at the
mesoscopic length scale $\ell\approx5\,$nm, but assumes that linear
elasticity does not hold at that scale. Indeed,
we do expect that linear elasticity will break down at length scales
corresponding to the curvature radius at which the DNA duplex is not a
minimum of free energy. We would not expect the
usual duplex form to be stable when bent into a loop of radius 5~nm.

The \SEC's other, less realistic features, such as the neglect of sequence
dependence, can readily be addressed, albeit at the cost of explicit
solutions. Its bending energy function, however, is not meant to be a
literal depiction of DNA mechanics. In principle, the true bending
energy function can eventually be deduced from statistical analysis of
sufficiently accurate determinations of DNA contours, obtained either
in solution via cryo EM, or when adsorbed to surfaces by AFM or EM. 
Alternatively, the short-length-scale bending energy might be  
calculated using molecular dynamics simulations. 
Direct all-atom molecular dynamics computations of the chain statistics for 
long-contour-length sequences of DNA are prohibitive computationally, 
but the generalized polymer model described in this paper is based upon 
the chain statistics of short-contour-length links which may be directly simulated.

\subsection{Future directions}

For many biological applications of DNA chain statistics, the twist degrees of freedom are also of great importance. For instance for DNA looping, moving an operator (the DNA binding sequence) a few base pairs can change the looping probability by an order of magnitude \cite{Muller1}.
This dramatic, short-contour-length dependence arises from the necessity of bring the DNA operator into twist registry with the binding site.
The twist degree of freedom of DNA has also been described by a
fluctuating elastic rod, the Helical Wormlike Chain model (HWLC)
\cite{YamakawaBook}. At long length scales, this modified WLC model
has successfully described the twist dependence of DNA. Nevertheless,
at high enough strain the HWLC model breaks down. For example, Bryant {\it et
al.} have demonstrated that the restoring torque
generated by twisted DNA saturates for high twist densities, implying
that the linear elastic model breaks down when the undertwist
$|\Delta\omega|$  exceeds 0.01~radian/basepair \cite{Bryant2003}. The
twist density needed to join a mis-phased DNA loop of under 100~bp
exceeds this threshold, and indeed Cloutier and Widom have also shown that the twist-induced
modulation of the cyclization $J$ factor is smaller for short sequences
than predicted by the HLWC model \cite{Cloutier2005}.

Thus, although the
bending of DNA for small twist densities may be adequately described
by the HWLC model, a generalized model of DNA, including elastic
breakdown of both bend and twist stiffnesses, may be
necessary to describe the chain statistics of short sequences of
looped DNA that are not naturally in twist registry when bound. Such
generalized models are in principle a straightforward extension of the
theory presented in this paper and new exact results for the HWLC
model recently derived by Spakowitz \cite{Andy3}.

\section{Conclusion}
In this paper, we have explored a class of generalized semiflexible polymer models in which the bending energy density is an arbitrary function of curvature. To analyze the chain statistics of these models, we developed a formalism that is analogous to the techniques used for describing the WLC model. We demonstrated that the statistics of these general models coincide with those of the linear-elastic (WLC) model at long contour length, as predicted by the renormalization group. At short length scales, we show that the predictions of these models can be dramatically different from the WLC model.  We computed near-exact expressions for the transformed spatial and tangent-spatial distribution functions with a method analogous to that recently exploited to find exact results for the WLC model. 
These generalized models provide an explicit example of a non-renormalizable model which is nearly exactly solvable. We exploited these general theoretical results to compute several important experimental observables: force-extension, the structure factor, and the cyclization $J$ factor. We explicitly performed these computations for a toy model of DNA bending, the Sub-Elastic Chain (SEC) model. The predictions of this model are essentially indistinguishable from the WLC model for force-extension, solution scattering, and long-contour-length cyclization measurements, despite  dramatic differences between the bending energies of the two models on short length scales. For short-contour-length cyclization experiments, general models generically predict large deviations from WLC behavior. In particular we computed the $J$ factor for the \SEC model and showed that this model could account for the anomalously large cyclization $J$ factor measured by Widom and Cloutier \cite{CW}. 
We expect these generalized models to be widely applicable for describing the high-curvature statistics of other semiflexible polymers.  

\begin{acknowledgments}
We thank
N. R. Dan, 
Cees Dekker, 
M. Inamdar,
Igor Kulic, 
Richard Lavery,
J. Maddocks
John Marko,
Fernando Moreno--Herrero, 
Rob Phillips, 
Prashant Purohit,
J. M. Schurr,
A. Spakowitz,
Thijn van der Heijden
R. James, Z.-G. Wang,
Jonathan Widom,
and 
Yongli Zhang
for helpful discussions and correspondence. 
PAW acknowledges grant support from an NSF graduate fellowship, the
Keck Foundation and NSF Grant CMS-0301657, and the NSF-funded Center
for Integrative Multiscale Modeling and Simulation. PN acknowledges
NSF Grant DMR-0404674 and the NSF-funded NSEC on Molecular Function at
the Nano/Bio Interface DMR04-25780.
\end{acknowledgments}



\appendix

\section{Explicit expressions for $g_l$}
\label{a:a}
It is straightforward to determine the $g_l$ eigenvalues of any propagator using the orthonormal eigenbasis of the angular momentum representation. In two dimensions, the $g_l$ are
\begin{equation}
\pnlabel{ggl2d}
g_l = \int_{-\pi}^\pi d\theta\ g(\vec{t}(\theta);\vec{e}_z)\,\exp il\theta,
\end{equation}
where $\theta$ is defined as the angle away from the $z$ axis: $\vec{t}(0) = \vec{e}_z$.
In three dimensions, the $g_l$ are
\begin{equation}
\pnlabel{ggl3d}
g_l = \int d^2\vec{t}\ g(\vec{t}(\theta);\vec{e}_z)\,P_l(\vec{t}\cdot\vec{e}_z),
\end{equation}
where the $P_l$ are the Legendre Polynomials and $\cos \theta = \vec{t}\cdot \vec{e}_z$.

\section{Stiff polymer limit}
\pnlabel{gstiffpol} In this section, we show that a narrowly distributed
fundamental tangent distribution function generically implies  WLC statistics at long contour length. In
dimension $D$, this calculation, though straight forward,  requires some technical
mathematics, but these technical details are not important for the interpretation of the result. 

We begin the derivation with the definition of the $l$th moment of
the tangent distribution function expressed in terms of the propagator \eref{ggl}
\begin{equation}
g_{l} = \left<l{\bf m}\right|{\cal G} \left|l{\bf m}\right>
\end{equation}
where rigid-body-rotational invariance implies that $g_l$ is independent of ${\bf m}$.
We insert two complete sets
of states into the tangent representation
\begin{equation}
g_{l} = \int d\vec{t}\,d\vec{t}{\,}'\ \left<l{\bf
m}\,|\vec{t}\,\right>\left<\vec{t}\,\right|{\cal
G}\left|\vec{t}{\,}'\right> \left<\vec{t}{\,}'|l{\bf m}\right>.
\end{equation}

We can now replace the matrix element of the propagator with the fundamental 
tangent distribution function $g(\vec{t};\vec{t}{\,}')$ (\eref{gcalG}). Remember that this function depends only on the relative deflection angle of the tangents. We therefore replace the integral over the second tangent with an integral over rotation matrices, ${\cal R}$, and make the substitution $\vec{t}{\,}'\equiv
{\cal R} \vec{t}$:
\begin{equation}
\pnlabel{gneedthis} g_{l} = \int d\vec{t}\,d{\cal R}\
\left|\frac{dt'}{d\cal R}\right|\left<l{\bf
m}\,|\vec{t}\,\right>g(\vec{t};{\cal
R}\vec{t}) \left<\vec{t}\,\right|{\cal D}^\dag_{\cal R}\left|l{\bf
m}\right>,
\end{equation}
where we represent the change in measure symbolically and 
we have introduced the rotation operator \cite{sakurai}
\begin{equation}
{\cal D}_{\cal R} \left|\vec{t}\, \right> \equiv \left| {\cal R} \vec{t}\, \right>.
\end{equation}

Our interest is in the case where the tangent distribution function is narrowly distributed. We
shall therefore  expand the rotation operator, $\cal D$, with respect to the rotation angles which we shall assume are small.  
The rotation operator can be expanded in terms of these angles and the rotation generators \cite{sakurai}
\begin{eqnarray}
{\cal D}_{\cal R} &=& \exp( -i\theta_{ij} {\cal L}_{ij})  \\
&=& 1-i\theta_{ij}{\cal L}_{ij}-{\textstyle
\frac{1}{2}}\theta_{ij}{\cal L}_{ij}\theta_{mn}{\cal L}_{mn}+...,
\end{eqnarray}
where the $\theta_{ij}=-\theta_{ji}$ are the components of the rotation angle which multiply the generators of rotations in the $ij$
plane. 

To evaluate the integral over the rotation matrices, we must now choose a set 
of $\theta$'s which give a single cover of the tangent space. 
Since $g(\vec{t};{\cal R} \vec{t}\,)$ is independent of $\vec{t}$, it is convenient to choose a coordinate system in which $\vec{t}$ is in the direction 
of the $D$ axis. (We shall return to the unrotated frame before performing the integral over $\vec{t}$.) 
In this new coordinate system, it is convenient to use the cover generated by the coordinates 
$\{{\theta}_{Di}\}_{1..D-1}$ while setting all other $\theta$'s to zero. 

We denote the average taken with respect to the distribution
function  by $\left<\ \right>$. 
Due to rigid-body-rotational invariance around the $D$ axis,
\begin{eqnarray}
\left<\theta_{iD}\right>&=&0 ,\\
\left<\theta_{iD}\theta_{nD}\right>&=&\left<\theta^2\right>\delta_{in}/(D-1),
\end{eqnarray}
where 
\begin{equation}
\theta^2\equiv \sum_{i=1}^{D-1}\theta_{iD}^2
\end{equation}
is the total deflection angle.

The nonzero matrix elements can be put in a coordinate invariant form
\begin{equation}
\left<lm\right|\left.\vec{e}_D\right>\left<\vec{e}_D\right|  {\cal L}_{Di}{\cal L}_{Di} \left| lm\right> = \left<lm\right|\left.\vec{e}_D\right>\left<\vec{e}_D\right|  {\cal L}^2 \left| lm\right>
 \end{equation}
since the added terms in the Casimir operator, ${\cal L}^2$, are zero on $\left|\vec{e}_D\right>$.  
We can now go back to the unrotated coordinate system by setting $\vec{e}_D=\vec{t}$. 

After integrating over the complete set of tangent vectors, the resulting moment is 
\begin{equation}
g_l = 1-{\textstyle
\frac{1}{2}}(D-1)^{-1}\left<\theta^2\right> \left<l{\bf m}\right|
{\cal L}^2 \left|l{\bf m}\right>+{\cal O}({\cal
L}^4\left<\theta^{4}\right>).
\end{equation}

Since this expression is only correct to ${\cal O}(\theta^4)$, it is 
convenient to replace ${\textstyle
\frac{1}{2}} \theta^2$ with $1-\cos\theta$.
We can now use the definition of the persistence length given in 
\eref{gpersistencelength} to eliminate the dependence on
$\left<\cos \theta\right>$:
\begin{equation}
g_l = 1-\frac{\ell}{2\xi}\left<l{\bf m}\right| {\cal L}^2
\left|l{\bf m}\right>+{\cal O}({\cal L}^4\ell^2/\xi^2).
\end{equation}
Finally, we reconstruct the propagator from its moments
\begin{equation}
\label{sple}
{\cal G} = \sum_{l,\bf m} g_l \left|l\, {\bf m}\right> \left<l\, {\bf m}\right| = 1-\frac{\ell}{2\xi} {\cal L}^2+{\cal O}({\cal L}^4\ell^2/\xi^2),
\end{equation}
which completes the derivation.
This result is discussed in \sref{gspersistence}.

\section{The transformed spatial propagator}
\pnlabel{gtsp}

%


To derive closed form expressions for the spatial propagator, we Fourier Transform the spatial propagator over the relative displacement, $\vec{X}$. In particular, we consider the Fourier Transform of \eref{gconvol} since in Fourier space, the spatial convolutions are simply products:
\begin{equation}
\pnlabel{gconvapp}
\tilde {\cal G}(\vec{k};L+L') = \tilde  {\cal G}(\vec{k};L)\tilde {\cal G}(\vec{k};L').
\end{equation}
We choose a coordinate system where $\vec{k}$ is in the $z$ direction.

We now wish to use this composition property of the spatial propagator to write a differential equation for ${\cal G}$. We therefore consider ${\cal G}$ for a differential arc length $dL$ and then expand the Fourier Transform of \eref{gsp} for arc length $dL$: 
\begin{equation}
\tilde {\cal G}(\vec{k};dL) = {\cal I}-{\cal A}dL,
\end{equation}
where ${\cal I}$ is the identity operator and ${\cal A } \equiv  {\cal H}+ik\cos \theta$ where $\theta$ takes its canonical meaning in spherical coordinates: $\cos \Theta = \vec{t}\cdot \hat{z}$. Substituting this expression into \eref{gconvapp}, we can write a differential equation for $\tilde{ \cal G}$:
\begin{equation}
\frac{d}{dL} \tilde {\cal G}(k;L) = - {\cal A} \tilde {\cal G}(k;L). 
\end{equation}
It is now convenient to make a Laplace transform from arc length $L$ to its conjugate variable $p$.  After solving for the propagation operator, we have an operator equation for the Laplace-Fourier Transform of the spatial propagator:
\begin{equation}
\pnlabel{gtransformedsp}
\tilde{\cal G}(k;p) = \{p{\cal I}+ {\cal A}(k)\}^{-1}=\{p{\cal I}+ {\cal H}+ik\cos \theta\}^{-1},
\end{equation}
but this expression is not explicit since it is written in terms of the inverse of an infinite dimensional operator.

We can express $\cos \Theta$ in the angular momentum basis. It is most convenient to define a set of ladder operators:
\begin{equation}
\cos \theta = a_++a_-,
\end{equation}
where the ladder operators are defined by
\begin{eqnarray}
a_+ &\equiv& \sum_{l=0}^{\infty} \sum_{m=-l}^l  A_{l+1,l,m} \left| l+1 \ m \right>\left<l\ m\right|, \\
a_- &\equiv& \sum_{l=0}^{\infty} \sum_{m=-l}^l  A_{l,l+1,m} \left| l\ m \right>\left<l+1\ m\right|, 
\end{eqnarray}
and the $A_{l,l+1,m}$ are:
\begin{equation}
A_{l,l+1,m} =A_{l+1,l,m} = \sqrt{\frac{\left(l-m+1\right)\left(l+m+1\right)}{\left(2l+1\right)\left(2L+3\right)}}.
\end{equation}
The ladder operators have the property that they increase (decrease) the total momentum quantum number of a state by plus (minus) one. 




Next, we obtain explicit expressions for the matrix elements of the transformed spatial propagator. The Hamiltonian is diagonal in the angular representation, so it is convenient to factor the spatial propagator (\eref{gtransformedsp}) into diagonal and nondiagonal factors:
\begin{equation}
\tilde{\cal G}(k;p) =\left[{\cal I}+ \{p{\cal I}+{\cal H}\}^{-1}ik(a_++a_-)\right]^{-1} \{p{\cal I}+ {\cal H}\}^{-1} ,
\end{equation}
and expand it in a power series
\begin{equation}
\pnlabel{gpowerseries}
\tilde{\cal G}(k;p) = \sum_{n=0}^{\infty} \left[ -ik \{p{\cal I}+{\cal H}\}^{-1}(a_++a_-)\right]^{n} \{p{\cal I}+ {\cal H}\}^{-1}.
\end{equation}

As a first step, we will compute a diagonal matrix element:
\begin{equation}
\tilde{G}_{lmlm} = \left<l\ m\right| \tilde{\cal G}(k;p) \left| l\ m\right>.
\end{equation}
Computing these matrix elements is achieved by grouping the infinite set of terms in \eref{gpowerseries} into sub sets which can be summed exactly \cite{Andy}. 

We introduce $\tilde{G}^+_{l'ml'm}$ which is the matrix element of a subset of the terms in \eref{gpowerseries}, in which there are only transitions to states with total momentum $l=l'$ or greater \cite{Andy}.  This matrix element can be defined recursively since only transitions to adjacent states are possible. The matrix element is the sum over $n$ of the matrix elements with $n$ transitions to and from the $l \ge l'+1$ states, which can be written in terms of $\tilde{G}^+_{l'+1,m,l'+1,m}$.
The terms of this matrix element, a geometric series, can be summed exactly  \cite{Andy2}:
\begin{eqnarray}
\tilde{G}^+_{lmlm} &=& \frac{1}{p+h_l}\sum_{n=0}^{\infty} \left[\frac{-k^2 A^2_{l,l+1,m} \tilde{G}^+_{l+1,m,l+1,m}}{p+h_l}\right]^n \\ &=& \left[p+h_l + k^2 A^2_{l,l+1,m} \tilde{G}^+_{l+1,m,l+1,m}\right]^{-1},
\end{eqnarray}
This sum is pictured schematically in \fref{ggplus}.

\pfign{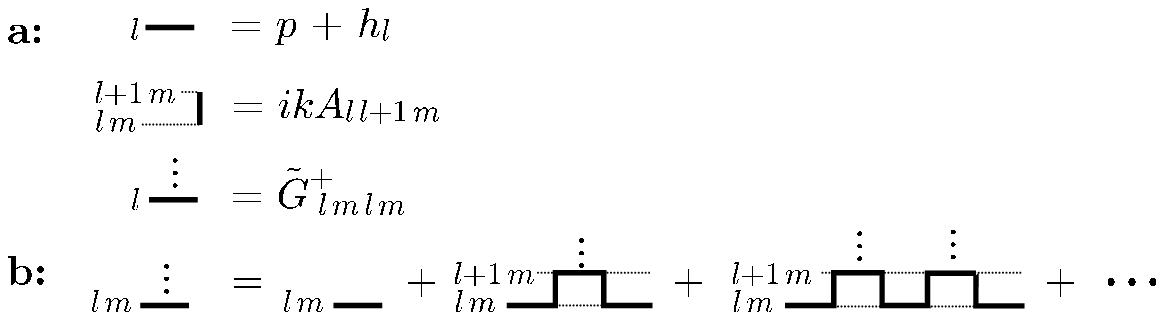}{ggplus}{Diagramatic rules for the propagator}{Diagramatic rules for the propagator: diagrams and their algebraic representations. Connected diagrams represent the products of the corresponding algebraic representations. The matrix element of the spatial propagator $\tilde{G}_{lml'm}$ is the sum of all diagrams which begin at state $l\,m$ and end at state $l'\,m$ with an arbitrary number of intermediate transitions.  (a) Horizontal lines represent propagation. Vertical lines represent transitions induced by the wave number.  $\tilde{G}^+_{lmlm}$ is the matrix element of the spatial propagator where transitions to states with total angular momentum $l-1$ or smaller are forbidden. This matrix element is represented by the line with ellipses, representing all transitions to states with higher $l$. (b) $G^{+}_{lmlm}$ can be defined recursively in terms of  $\tilde{G}^+_{l+1,m,l+1,m}$. The definition of $\tilde{G}^-_{lmlm}$ is analogous, but it is the sum of all diagrams with transitions to states with total angular quantum number $l$ and smaller.}

Similarly, we define $\tilde{G}^-_{l'ml'm}$ which is the matrix element of the propagation operator which allows transitions to states with total momentum $l=l'$ or less: 
\begin{equation}
\tilde{G}^-_{lmlm} = \left[p+h_l + k^2 A^2_{l,l-1,m} \tilde{G}^-_{l-1,m,l-1,m}\right]^{-1}.
\end{equation}
In terms of $G^\pm$ we can now define the matrix element without transition restrictions by grouping the transitions into sets that do not cross $l=l'$. These sets can be written in terms of  the matrix elements of ${\cal G}^\pm$ and then summed in a geometric series \cite{Andy2}:
\begin{equation}
\pnlabel{gpropmatrixelement}
\tilde{G}_{lmlm} = \left[p+h_l + k^2 A^2_{l,l+1,m} \tilde{G}^+_{l+1,m,l+1,m}+ k^2 A^2_{l,l-1,m} \tilde{G}^-_{l-1,m,l-1,m}\right]^{-1}.
\end{equation}
The diagonal matrix element computed above is sufficient for describing many observables of phenomenological interest. Note that the only difference between this expression and the WLC expression \cite{Andy2} is that the eigenvalues of the Hamiltonian operator have changed.

For some applications we will want completely general matrix elements $\tilde{G}_{lml'm'}$. We can again define these general matrix elements in terms of the recursive definitions of $G^\pm$. Again, the trick to summing the terms is grouping them. In this general case, there are many equivalent ways of achieving this grouping. See \fref{gnotequal} for an explanation of the set grouping. The matrix element can be written \cite{Andy2}
\begin{equation}
\tilde{G}_{l+n,m,l,m'} = \tilde{G}_{l,m,l+n,m'} = \delta_{m-m'}\tilde{G}_{lmlm} \prod_{q=1}^{n}  -ik A_{l+q-1,l+q,m} \tilde{G}^+_{l+q,m,l+q,m}.
\end{equation}
We have now explicitly solved for spatial propagator having written expressions for all the matrix elements.

\pfign{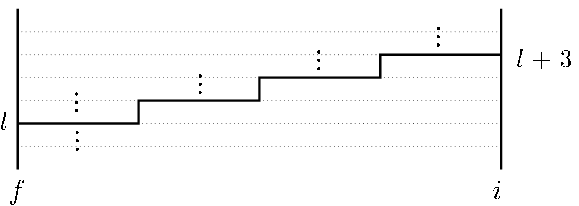}{gnotequal}{General matrix elements}{General matrix elements.  A diagram of the sum for the matrix element $\tilde{G}_{lml+nm}=\tilde{G}_{l+nmlm}$. To compute the matrix element, we group the terms by the location of the first steps from $l+n$ to $l+n-1$ and from $l+n-1$ to $l+n-2$ etc. In the diagram, these steps are represented by the vertical lines. We use the ${\cal G}^+$ operator to sum over all possible diagrams with upward transitions between these steps. These upward transitions are represented by the ellipses. We multiply by the transition matrix element for each of the vertical lines. After we reach $l$ for the first time we allow all transitions up or down. This enumeration counts each contributing diagram once but this recipe is not unique.}

\section{The computation of spatial distributions}
\label{numericalsd}
The previous section discussed near-exact expressions for the
Fourier-Laplace transformed spatial and tangent-spatial
distribution functions. Exact closed-form expressions for 
these functions are unknown and we must invert the transforms numerically 
to compute the distribution functions. 

\subsection{Force-extension and the structure factor}
The computations of force extension and the structure factor require only a single numerical inverse Laplace Transform.  We cut off the continued fraction at $l=10$ and then used the {\tt InverseLaplaceTransform} function in {\it Mathematica}.

\subsection{The spatial distribution and the $J$ factor}

For computations of the spatial distribution function and the $J$ factor, we exploited two different numerical techniques: numerical transform inversion and Monte Carlo. For contour lengths of a persistence length and above, it is convenient to directly invert the transforms numerically by truncated the continued fraction in the transformed propagator (\eref{gpropmatrixelement}). 
Typically we used $l=15$ as the cutoff although in some cases higher $l$ values were used for short
contour lengths.

In the inverse transform technique, both numerical Laplace and Fourier Transform inversions must be computed. We have used two different implimentations for these computations. {\it (i)} 
In {\it Mathematica}, we used the {\tt InverseLaplaceTransform} function.  We then integrated numerically (using an explicit sum) to invert the Fourier transform. We found that the built-in numerical integration in {\it Mathematica} was too slow for practical use. 
{\it (ii)} In {\it Matlab}, we used a code which explicitly computed the Laplace Transform by computing the sum of the residues of the inverse Laplace Transform contour integral. The Fourier Transform inversion was again performed by numerical integration using an explicit sum. 
The {\it Matlab} code was based on one shared with us by Andy Spakowitz.

For contour lengths on order a persistence length and
shorter, inverting the transformed expressions is impractical. 
The continued fraction in increasing momentum 
is essentially an expansion around weak end-tangent correlation. 
For contour lengths shorter than a persistence length, a larger $l$ cutoff is
required, significantly slowing the numerical
inversions. In addition, the numerical integration over the wave number 
becomes impractical since the numerical integrations must be extended to very 
a large cutoff momentum.
These convergence
issues are not unique to the continued fraction approach. For
example, the transfer matrix approach is plagued by similar
shortcomings, requiring difficult numerical work
at short contour length \cite{Yan2004}.

We therefore used a much simpler,
although less elegant, solution in the form of direct Monte Carlo
integrations. Monte Carlo integration in the short-contour-length
regime {\it (i)} is numerically more efficient than direct inversion,
{\it (ii)} requires very minimal implementation, and {\it (iii)} serves as a useful check
of our theoretical results. These checks appear few places explicitly
in the paper since the agreement between these two methods is
excellent and the focus of this paper is physics rather than
numerical computations. The theoretical curve for the 
cyclization $J$ factor (\fref{gcycl}) contain both inversion and Monte
Carlo computations. 


%

%

\end{document}